\documentclass[%
aip,
 amsmath,amssymb,
 reprint,%
]{revtex4-1}

\usepackage{graphicx}
\usepackage{dcolumn}
\usepackage{bm}

\usepackage[utf8]{inputenc}
\usepackage[T1]{fontenc}
\usepackage{mathptmx}
\usepackage{etoolbox}
\usepackage{xcolor}
\usepackage{lipsum}
\usepackage{listings}

\newcommand{\DIPC}[0]{
Donostia International Physics Center (DIPC),
Paseo Manuel de Lardizabal 4, 20018 Donostia-San Sebasti\'an, Spain}
\newcommand{\UPV}[0]{Kimika Fakultatea, Euskal Herriko Unibertsitatea EHU, Donostia-San Sebastián 20018, Spain}

\newcommand{\BDX}[0]{Institut des Sciences Moléculaires (ISM), Université de Bordeaux, 351 Cours de la Libération, 33405 Talence, France}

\newcommand{\PSC}[0]{Universit\'e Paris-Saclay, CNRS, Institut des Sciences Moléculaires d’Orsay, 91405 Orsay, France}

\newcommand{\ULO}[0]{CNRS-Université de Lorraine, LPCT, 57070 Metz, France}

\begin{document}

\preprint{AIP/123-QED}
\title[]{Quantum Dynamical and isotopic effects for Hydrogen isotopes scattering at W(110) surface}
\author{Raúl Bombín}
\email{raul.bombin@u-bordeaux.fr}
\affiliation{\BDX}
\affiliation{\DIPC}

\author{Oihana Galparsoro}
\affiliation{\UPV}

\author{Daniel Pel\'aez}
\affiliation{\PSC}

\author{Jean Christophe Tremblay}
\affiliation{\ULO}

\author{C\'edric Crespos}
\affiliation{\BDX}
\author{Pascal Larregaray}
\affiliation{\BDX}

\date{\today}

\begin{abstract}
We investigate the scattering of hydrogen isotopes at the W(110) surface using both classical and quantum dynamics approaches to elucidate the role of quantum effects in this system. 
To characterize the scattering process we focus on key observables, including the absorption probability and diffraction channels that we evaluate at the quasi-classical and quantum levels. 
The quantum dynamical simulations reveal pronounced resonance structures in the absorption curve that we rationalize in terms of diffraction-mediated selective adsorption and focused sticking mechanisms. Diffraction probabilities for reflected trajectories exhibit strong quantum effects at low incident energies, where classical dynamics underestimate the back scattering probability. 
These effects become less pronounced with increasing isotope mass, from hydrogen to tritium, but discrepancies between the classical and quantum descriptions persist at low incident energies.
\end{abstract}

\maketitle

\section{\label{sec:intro} Introduction}

The interaction of atoms and molecules with solid surfaces has been a paradigmatic problem in chemical physics and surface science. Over the last decades, developments in electronic-structure calculations and dynamical simulation techniques have made it possible to investigate these systems with increasing accuracy. Such studies provide a microscopic perspective on elementary gas--surface processes and offer a framework for examining how theoretical models relate to experimentally observable phenomena.\cite{Auerbach2021}

An important question in this context concerns the level of theory that is required to accurately describe elementary gas--surface interactions.
There are at least three research directions toward achieving this objective. First, substantial progress has recently been achieved through the development of highly accurate potential energy surfaces (PESs) based on first-principles calculations using machine-learning (ML) techniques.\cite{Zhou2021} These advances have enabled increasingly realistic simulations of dynamical processes at surfaces for high dimensional systems. Second, for light projectiles interacting with metal surfaces, not only phonon excitations but also non-adiabatic energy coupling due to electron-hole pair (ehp) (de-)excitations play a significant role. This highlights the necessity  to go beyond of the Born-Oppenheimer approximation. Third, although classical dynamics has revealed itself as a powerful tool to study gas--surface processes, at low energies, quantum effects may be important and thus a quantum mechanical treatment of the nuclear motion is desirable.

Beyond their fundamental interest, gas-surface processes are important for technological developments such as those in nuclear fusion research. Tungsten is the prime candidate material for plasma-facing components of future fusion reactors such as ITER,~\cite{Campbell2024_ITER_IRP} primarily because of its high melting point and favorable thermo-mechanical properties. Understanding hydrogen isotope retention and energy dissipation mechanisms at tungsten surfaces is therefore essential for predicting fuel recycling and material degradation under reactor operating conditions. Large-scale experimental programs, such as WEST and JET, are actively investigating the performance of plasma-facing components under deuterium and tritium operation.\cite{Bucalossi2024,Gallo2024,Maggi2024} These efforts are complemented by \textit{ab-initio} studies addressing deuterium trapping at tungsten dislocations~\cite{Jin2024} and fuel retention at high energies.~\cite{Hodille2025}

Hydrogen isotopes interacting with tungsten surfaces constitute an especially relevant model system, as due to their small mass, quantum effects are expected to be enhanced.
In particular, the combined fundamental and technological relevance of the H/W system has motivated extensive fundamental research aimed at understanding its microscopic dynamics both in molecular and atomic form. Several studies based on chemically accurate PESs have established a detailed description of H$_2$ interacting with tungsten surfaces.~\cite{Busnengo2008,Galparsoro2016,Galparsoro2018,Rodriguez2021,Viaud2024,Viaud2025} Subsequent investigations of atomic scattering demonstrated the importance of non-adiabatic effects, particularly energy dissipation through ehp (de-)excitations in the metal. These effects have been modeled using Langevin dynamics that incorporate frictional and stochastic forces,~\cite{Hertl2021,Barrios2022} successfully reproducing experimental observations for hydrogen scattering from metal surfaces.~\cite{Bunermann2015_Science,Ferrin2012,Dorenkamp2018} Extensions of these studies examined the influence of hydrogen coverage and incidence angle on energy-loss distributions,~\cite{Barrios2021,Barrios2024} while alternative descriptions based on effective-medium theory has also been applied to tungsten and other bcc metals.~\cite{Hertl2022} The dissipation of energy due phonon excitations have also been studied by accounting for the surface degrees of freedom, whether by performing \textit{ab-initio} molecular dynamics calculations for tungsten covered surfaces~\cite{Rodriguez2023,Omar2025} or, more recently, using multidimensional ML-PESs,~\cite{Stark2023,Yang2025} what has enabled investigations of isotopic and coverage effects in the H/W system.~\cite{Yang2025}

Despite these advances, most dynamical studies of hydrogen scattering from tungsten surfaces have been performed within a classical framework. While such approaches can successfully account for several experimental observables, they do not explicitly address the role of nuclear quantum effects, including phenomena such as selective adsorption resonances (SARs). A natural first step toward investigating these effects is to consider a frozen-surface Born--Oppenheimer static surface (BOSS) description, in which energy dissipation through phonons and electron-hole pair excitations is neglected. Although these dissipation channels are known to play a significant role in H/W(110) dynamics,\cite{Hertl2021,Barrios2021} such a model provides a well-defined framework for isolating quantum nuclear effects. Resonance features associated with bound states supported by the corrugated PES are expected to remain identifiable, although broadened and shortened in lifetime, when dissipation is included. This reference description, therefore, provides a basis for distinguishing these elastic resonances from additional resonance features that will emerge when non-elastic processes are explicitly incorporated.

Quantum dynamical treatments of gas--surface processes have a long history. Early time-dependent wave-packet (TDWP) studies employed reduced-dimensionality models to investigate hydrogen recombination on graphite~\cite{Meijer2001} and H$_2$ scattering from Cu(111)~\cite{Persson1995,Jackson1995} and Pt(111)~\cite{Pijper2000}. Subsequent developments enabled six-dimensional quantum descriptions of molecular scattering and dissociation on metal surfaces.~\cite{Kroes1999,Pijper2001,Diaz2005,Diaz2010} More recently, the Eley--Rideal recombination of hydrogen on Cu has been studied using TDWP on a six-dimensional PES~\cite{Xiong2024,Xiong2024JPCC} for the first time.
Regarding atomic scattering at surfaces, that has been proposed as a reference system for benchmarking theoretical models with accurate experimental measurements,~\cite{Bunermann2021_JPCA} hydrogen scattering from graphene has been investigated using the multi-configurational time-dependent Hartree (MCTDH) method~\cite{Shi2023,Shi2025}, complementing earlier studies on this system.~\cite{Sha2002,Bonfanti2015,Bonfanti2018,Jiang2021} For metallic systems, Lepetit \textit{et al.}~\cite{Lepetit2011} examined low-energy scattering in the van der Waals well of graphite using TDWP methods. 
MCTDH has been applied to the grazing incident fast-atom diffraction on semiconducting surfaces (GIFAD) of He Li(0001)~\cite{Gravielle2014,Muzas2024} and KCl,~\cite{Bocan2020} showing reasonable good agreement with exact wave-packet results. Note, however,
that these works consider projectile energies in the range of kiloelectronvolts, that are much larger than those considered in works motivated by catalysis.
Nevertheless, to the best of our knowledge, no quantum dynamical studies have addressed hydrogen scattering from metal surfaces in the energy regime where the projectile probes the chemisorption well and strong gas--surface interactions dominate the dynamics. 

The limited number of quantum studies of atomic scattering from metal surfaces leaves the role of nuclear quantum effects in these systems largely unexplored. Here, we investigate H scattering from W(110) using quantum TDWP calculations within the MCTDH framework. By comparing quantum and classical dynamics on the same PES, we assess the impact of quantum nuclear motion on scattering observables and on the theoretical description of hydrogen–metal interactions. As a well-established approach for treating systems with multiple degrees of freedom, MCTDH provides a promising route for future gas–surface dynamics studies that explicitly account for surface motion.

The paper is organized as follows. In Section~\ref{sec:Method}, we detail the methodology used to conduct the classical and quantum dynamical simulations. Results are presented in Section~\ref{sec:results}, and finally the conclusions are summarized in Section~\ref{sec:conclusions}. In Appendix~\ref{app:vib}, the vibrational spectra of hydrogen isotopes are discussed. In Appendix~\ref{app:convvib} we discuss the convergence of the calculated vibrational spectra.

\begin{figure}
\includegraphics[width=1.0\linewidth]{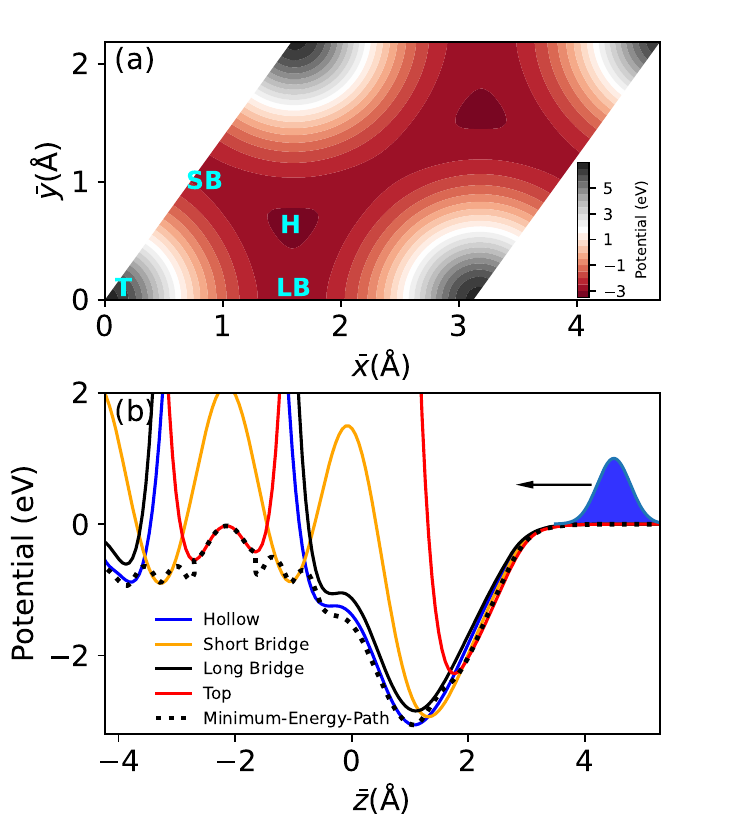}
\caption{\label{fig:potZcuts} (a) 2D PES cut for H/W(110) system at a distance ${\rm Z}1.1$~{\AA} from the surface. Four high-symmetry positions are labeled: Hollow (H), short-bridge (SB), long-bridge (LB) and top (T) sites. (b) $Z$-cuts of the PES at the high-symmetry positions.
Here $\bar{x}$, $\bar{y}$ and $\bar{z}$ are orthogonal cartesian coordinates.
The potential energy along the minimum energy path is shown in dotted black line.}
\end{figure}

\section{\label{sec:Method} Method}

\subsection{\label{subsec:PES} Potential Energy Surface}
Classical trajectories (CT) and TDWP calculations are performed by propagating the system on a PES. 
We make use of the Born–Oppenheimer static surface (BOSS) model, in which energy dissipation to the surface vibrations is neglected. 
Despite the large mismatch between H and W masses (m$_H$/m$_W$ =1/183.5), the kinetic energy that the projectile acquires when it approaches the surface can make phonon excitations important, as revealed by several authors doing classical dynamics simulations with different techniques, including the employment of modern ML-PES (see for example Refs.~\onlinecite{Becerra2020,Stark2023,Yang2025}). The latter effect is expected to be enhanced for the heavier isotopes of H.
Our model also intentionally neglects energy dissipation to the surface electrons. Its theoretical treatment within the quantum framework remains challenging~\cite{Bi2024,Bridge2024,Chen2025,Preston2025} and is currently the subject of intense research. We defer this aspect to future work. The objective of the present study is to evaluate the importance of quantum effects in H scattering on a W(110) model surface.

The PES~\cite{Galparsoro2016,Petuya2014} is based on density functional theory  calculations performed with the PW91 exchange correlation functional that relies on the generalized gradient approximation. 
To interpolate the PES, we employ the corrugation-reducing procedure (CRP),~\cite{Busnengo2000,Olsen2002,Busnengo2008} which has been widely applied in studies of atomic and diatomic molecular scattering on various metal surfaces.~\cite{Kroes1999}
To propagate the TDWP on the PES, we use the MCTDH method, as implemented in the Heidelberg package,~\cite{Beck2000} which reduces the computational cost of quantum calculations by representing the wave function as a sum of Hartree products of nuclear orbitals.
To fully exploit this approximation and maximize computational efficiency, the PES must also be expressed in separable form. A sum-of-products representation is obtained using the POTFIT program\cite{Jackle1996} included in the Heidelberg package. Further details about the POTFIT representation are given in section~\ref{sec:method-Q}.

Although the CRP method is able to describe a 6D potential for a diatom interaction with the surface, here we consider single atom scattering, thus the potential is a three dimensional (3D) function of the position of the atom, $V^{\rm 3D}(x,y,z)$. 
In the choice of coordinates, the ${XY}$ plane is parallel to the surface and the $Z$ direction is perpendicular to it, with ${\rm Z}0$ corresponding to the height of the topmost layer of W atoms. The $X$ and $Y$ axes coincide with the W(110) surface's cell skewed lattice vectors, that form an angle $\gamma =$~54.735$^\circ$ between them, being the lattice constant 3.17~\AA. The symmetry of the surface is illustrated in Fig~\ref{fig:potZcuts}(a) by a PES cut in the ${\rm XY}$ plane. 
The height of the cut has been chosen to coincide with the equilibrium position of an H atom adsorbed on the tungsten surface (${\rm Z}1.1$~{\AA}), that is, at a hollow site. 
The black regions represent the repulsive barriers that exist on tungsten atop positions at this height, while the adsorption site is signaled by the dark red region, being the adsorption energy $E_{\rm ads}\sim-3.1$~eV. 
Fig.~\ref{fig:potZcuts}(b) depicts one-dimensional cuts of the PES along $Z$-direction for the high-symmetry positions that are labeled in panel (a): hollow (H), short-bridge (SB), long-bridge (LB) and atop (T) sites, all of them exhibiting local energy minima with increasing energy. Notice, however, that the latter one does not correspond to a local minima on the 3D PES but to a saddle point. In addition, the potential energy cut along the minimum energy path for absorption, calculated using the nudged elastic band method,\cite{Henkelman2000} is represented by the dashed curve. As it can be seen, our PES is able to describe the system for gas phase configurations ($z>$ 5\AA) as well as bulk penetration up to a depth of ($-4.5$~\AA).  As atoms arrive to the surface from the gas phase, they encounter barriers at the different high-symmetry sites, lying in the range $z\in[-2.0, 1.5]$\AA, from where they are either reflected or deflected. 
For small values of the incident energy $E_{\rm in}$, in order to penetrate to the bulk region, an hydrogen atom should first penetrate the surface through hollow or long-bridge positions and then follow a trajectory over short bridge, top and long bridge (hollow), iteratively. A minium energy path, following such type of trajectory in depicted with dotted black line is Fig.~\ref{fig:potZcuts}(b).

\subsection{\label{subsec:clastraj}Classical trajectory simulations}
\label{sec:method-C}

In the CT calculations, the  nuclear motion of the hydrogen isotope is described by solving Newton’s second law,

\begin{equation}
    m_i \ddot{\mathbf{r}}_i = -\frac{d V^{\mathrm{3D}}}{d \mathbf{r}_i},
    \label{eqn:newton}
\end{equation}
where $V^{\mathrm{3D}}$ is the adiabatic PES, and $m_i$, $\mathbf{r}_i$, and $\ddot{\mathbf{r}}_i$ denote the mass, position, and acceleration of the isotope $i = \mathrm{H}, \mathrm{D}, \mathrm{T}$.

Normal incidence scattering is simulated by randomly sampling the initial in-plane coordinates of the projectile atom $(x,y)$, taking advantage of the symmetry of the surface. The initial altitude of the projectile is $z_0$ =7.0~{\AA}, which corresponds to the asymptotic region of the atom-substrate interaction potential. 
Our model does not consider energy dissipation to the surface degrees of freedom, thus, there are two possible exit channels for the simulations: absorption in the bulk and reflection. An atom is considered reflected when it reaches its initial height $z_0$ and absorbed when it crosses the plane located at ${\rm Z}-2.5$~\AA. 
Notice that for short propagation times, dynamical trapping occurs at the surface. 
For the results presented here, trajectories are propagated to long enough times so that the dynamical trapping probability vanishes.
At the lowest incident energy considered ($E_{\rm in }=$~10 meV), propagation times of up to 20 ps are required.
To reach convergence for the absorption curves, 1000 trajectories are performed for each value of the incident energy, whereas converged diffraction patterns require up to 8 million trajectories per incident energy. The larger statistical sampling required for the latter arises from the Gaussian projection of the classical trajectories onto the discrete quantum diffraction channels, together with the final-state-resolved nature of the observable. Convergence was verified with respect to the Gaussian width and the number of trajectories.

\subsection{\label{subsec:mctdh}Wave-packet simulations}
\label{sec:method-Q}

The state of the system is represented as a wave-packet, $\Psi(X,Y,Z;t)$,
that evolves according to the time-dependent Schr\"odinger equation
\begin{equation}\label{tdse}
    i\hbar\frac{\partial\Psi(X,Y,Z;t)}{\partial t} = \hat{H}\Psi(X,Y,Z;t)
\end{equation}
The Hamiltonian, $\hat{H}$, describing the interaction of an atom of mass $M$ with a frozen surface is given by
\begin{eqnarray}
\hat{H} &= -\frac{\hbar^2}{2M\sin^2\gamma}\left[\frac{\partial^2}{\partial^2 X}+\frac{\partial^2}{\partial^2 Y}-2\cos\gamma\frac{\partial}{\partial X}\frac{\partial}{\partial Y}\right] \nonumber\\
&-\frac{\hbar^2}{2M}\frac{\partial^2}{\partial^2 Z} + V^{\rm 3D}(X,Y,Z)~,
\label{eqn:hamiltonian}
\end{eqnarray}
where we have employed non-orthogonal, skewed coordinates $\{X,Y,Z\}$, which are related via the skewing angle $\gamma = 54.735^\circ$ to the Cartesian ones, $\{\bar{x},\bar{y},\bar{z}\}$, by the relations
\begin{equation}
X = \bar{x} -\bar{y}/\tan(\gamma)\quad;\quad 
Y = \bar{y}/\sin(\gamma)\quad;\quad 
Z =\bar{z}~.
\end{equation}

To perform the TDWP calculations, we rely on the MCTDH method as implemented in the Heidelberg package.~\cite{Beck2000} The wave function that describes the time-evolution of the systems is expanded as a sum of Hartree products of single particle functions (SPFs), that in our 3D case reads
\begin{equation}\begin{aligned}
    \Psi(&X,Y,Z;t) =\\ &\sum_{j_x=1}^{n_x}\sum_{j_y=1}^{n_y}\sum_{j_z=1}^{n_z} A_{j_xj_yj_z}(t) \varphi_{j_x}(X,t)\varphi_{j_y}(Y,t)\varphi_{j_z}(Z,t)~.
    \label{eqn:hartreeWF}
\end{aligned}\end{equation}

Here, $A_{j_xj_yj_z}(t)$ are time-dependent expansion coefficients.
A number $n_i$ of SPFs $\{\varphi_{j_i}\}$ is chosen to represent accurately the $Q_\kappa=\{X,Y,Z\}$ nuclear degree of freedom.
In the Heidelberg implementation of MCTDH, the SPFs
for the generalized $Q_\kappa$ coordinates are further expanded in a time-independent primitive basis of functions $\{\chi_{i_\kappa}(Q_{\kappa})\}$ as
\begin{eqnarray}
    \varphi_{j_\kappa}( Q_\kappa,t) = \sum_{i_\kappa=1}^{N_\kappa}c_{i_\kappa j_\kappa}(t)\chi_{i_\kappa}(Q_{\kappa})~,
    \label{eqn:SPFexpansion}
\end{eqnarray}
with $c_{i_\kappa j_\kappa}(t)$ the time-dependent coefficients of the SPF and $N_\kappa$ the number of grid points associated to the $\kappa$-th degree of freedom.
Substituting Eq.\,\eqref{eqn:hartreeWF} into Eq.\,\eqref{tdse} and applying the Dirac-Frenkel variational principle leads to coupled equations of motion (EOM) for the coefficients, $A_{j_xj_yj_z}(t)$, and the SPFs, via the evolution of the contraction coefficients $c_{i_\kappa j_\kappa}(t)$. These EOM are solved numerically using various integration schemes, as described in detail elsewhere. \cite{Beck2000}
Here, we make use of the so-called constant mean-field integration (CMF) scheme (relative tolerance: 10$^{-5}$), using the Burlisch-Stoer algorithm (steps: 7, relative tolerance: 10$^{-7}$) and Short-Iterative Lanczos algorithm  (iteration depth: 12, relative tolerance: 10$^{-6}$) for the propagation of the SPFs and coefficients, respectively.
\begin{table}[tb]
    \caption{Number of SPFs, $n_\kappa$, per degree of freedom, $Q_\kappa = \{X, Y, Z\}$,  for the MCTDH calculations used in this work for the simulation of H, D and T scattering on the W(110) surface.}
    \label{table:SPFs}
    \begin{ruledtabular}
    \begin{tabular}{c ccc }
        isotope & $n_x$ & $n_y$ & $n_z$ \\ \hline
        H & 50 & 50  & 100 \\
        D & 61 & 54 & 105 \\
        T&  65&  56 & 110 
    \end{tabular}
    \end{ruledtabular}
\end{table}
Although the Hartree wave function of Eq.~\eqref{eqn:hartreeWF} does not account exactly for nuclear quantum correlation effects, the multi-configurational nature of the MCTDH ansatz accounts for dynamical correlation between the different degrees of freedom, induced by the corrugation of the PES. 
To correctly describe this nuclear correlation, 
the convergence of the SPF basis with respect to $\{n_x, n_y, n_z\}$ was carefully checked. 
Table~\ref{table:SPFs} summarizes the parameters used in the calculations for the different isotopes.
Note that a larger number of SPFs is required to obtain converged results for heavier isotopes. 
At first glance, this may seem contradictory with the observation that quantum effects are less pronounced in heavier atoms.
Nevertheless, it can be understood from the fact that the same potential wells support more bound states for the heavier isotopes.
Consequently, a larger basis is required to describe the intermediate scattering states for a given value of the incident energy.
As a result, the computational cost increases sharply with increasing isotope mass.

Throughout this work, we use the Fast Fourier Transform periodic collocation method for the primitive basis.
The SPFs are represented using $N_x=67$ points on the interval $[0,3.170]$\,\AA{}, $N_y=58$
points on the interval $[0,2.745]$\,\AA{}, and
 $N_z=316$ points on the interval $[-4.445,7.832]$\,\AA{}.
The grid spacing, $\Delta Q_\kappa$, is 4.7$\times 10^{-2}$\AA{} along $X$ and $Y$, and 3.9$\times 10^{-2}${\AA} for $Z$. 
This spacing is sufficient to resolve the maximum momentum ($k_{\max}^{Q_\kappa}= \hbar/\Delta Q_\kappa$) attained by the particle upon entering the surface potential wells, The corresponding kinetic energy is limited by $E_{\rm in}$ and the depth of the potential, $E_k \approx E_{\rm in} + 3.1$~eV.
Despite of the this consideration, the convergence of our calculations is checked in terms of grid spacing between the primitive basis functions (see appendix~\ref{app:convvib}, where the convergence of the vibrational spectra of H at the W(110) surface is shown.).
As required by MCTDH, in all simulations, the PES has been represented on the primitive grid in the so-called sum-of-products form using the POTFIT program.~\cite{Jackle1996} The sum of products is constructed using 63, 51  natural potentials for the $X$, $Y$ degrees of freedom, and $316$ (equal to $N_z$) for $Z$. This yields a root mean square  error of 0.2~meV, while the maximum error remains below 4~meV. This difference in the number of potential basis functions responds to the unbound nature of the last degree of freedom.

The initial wave-packet in all of the simulations has been prepared as a Hartree product of constant functions ($k=0$ plane waves) for the $X$ and $Y$ degrees of freedom and a complex-valued Gaussian function for the $Z$ one. The initial width of the Gaussian is chosen as $\sigma=0.42$\,\AA{} and centered at $Z_{\rm 0}=4.76$\,\AA{}, with variable initial momentum $p_0$ defined from the incident energy $E_{\rm in}$ of the particle in gas phase.
To avoid artificial reflections on the borders of the box along the $Z$ direction, two complex absorbing potentials (CAP) have been placed at $\rm Z^{\rm CAP}_{\rm ref}=5$~{\AA} and $\rm Z^{\rm CAP}_{\rm abs}=-2.5$~{\AA} for the reflection and the absorption channels, respectively. The form of the CAP is taken as $V^{CAP}=-iV_0^{CAP}\theta\big(\pm(z-z^{\rm CAP})\big)\big(z-z^{\rm CAP}\big)^2$ with the plus (minus) minus sign for the reflection (absorption) channel. 
The values $V_0^{CAP}$ are chosen for each initial incident energy so that they minimize spurious reflection effects while preventing to have a significant population in the first (last) primitive basis point in the $Z$-direction for the absorption (reflection) CAP.
The isotope and $E_{\rm in}$ dependent values of $V_0^{CAP}$ are listed in Table~\ref{table:CAP}. Note that the initial wave-packet has a finite overlap with the region where the reflection CAP is placed. To prevent artificial absorption at the beginning of the propagation, the CAP is activated only after the overlap has become negligible and before the TDWP re-enters the reflection CAP region.
\begin{table}[tb]
\caption{$V_0^{\rm CAP}$ values used for MCTDH simulations for H, D, and T depending on their initial incident energy, $E_{\rm in}$. 
For a wave-packet with initial energy $E_{\rm in}$ lying within one of the intervals listed in the H, D, or T column, the corresponding $V_0^{\rm CAP}$ value is used.}
\label{table:CAP}
    \begin{ruledtabular}
    \begin{tabular}{cccc}
     & \multicolumn{3}{c}{$E_{\rm in}$ range (eV)}  \\\hline
     $V_0^{\rm CAP} (\times 10 ^{-4}$ a.u.) & H & D & T  \\\hline
     $0.8$ & [0.01,0.07) & [0.01, 0.10) &  [0.01, 0.23) \\
     $2$ & [0.07, 0.75) & [0.10, 0.60) & [0.23, 0.90]\\
     $4$ & [0.75, 1.10] & [0.60, 1.10] \\
    \end{tabular}
    \end{ruledtabular}
\end{table}
To analyze the results, we compute the flux that crosses the CAPs, which gives access to absorption and reflection probabilities. To obtain the diffraction patterns, the flux that crosses the reflection channel is projected onto the diffraction states.
For state-resolved analyses, bound vibrational states have been obtained from imaginary time propagation of an initial wave-packet localized in the vicinity of the surface. To this end, we employ the improved relaxation algorithm,\cite{mey06:179} using a CMF integration scheme with an 8th order Runge-Kutta algorithm for the propagation of the SPFs and a Davidson diagonalization for the A-coefficients.

Although the present work is limited to a three-dimensional adiabatic model, MCTDH provides a promising framework for extending quantum dynamical studies of gas--surface interactions to more realistic descriptions that explicitly account for the relevant energy dissipation channels at the surface. In its multilayer formulation,~\cite{Song2022} MCTDH can treat systems with many degrees of freedom, enabling the explicit inclusion of surface motion.~\cite{Shi2025} Furthermore, the recently developed stochastic formulation of MCTDH can describe systems with non-adiabatic couplings,~\cite{Mandal2022} thereby opening a route toward incorporating energy dissipation arising from electron-hole pair (de-)excitations.

\section{\label{sec:results} Results}
\subsection{Absorption curve}
\label{sec:res-abs}
\begin{figure}
\includegraphics[width=1.0\linewidth]{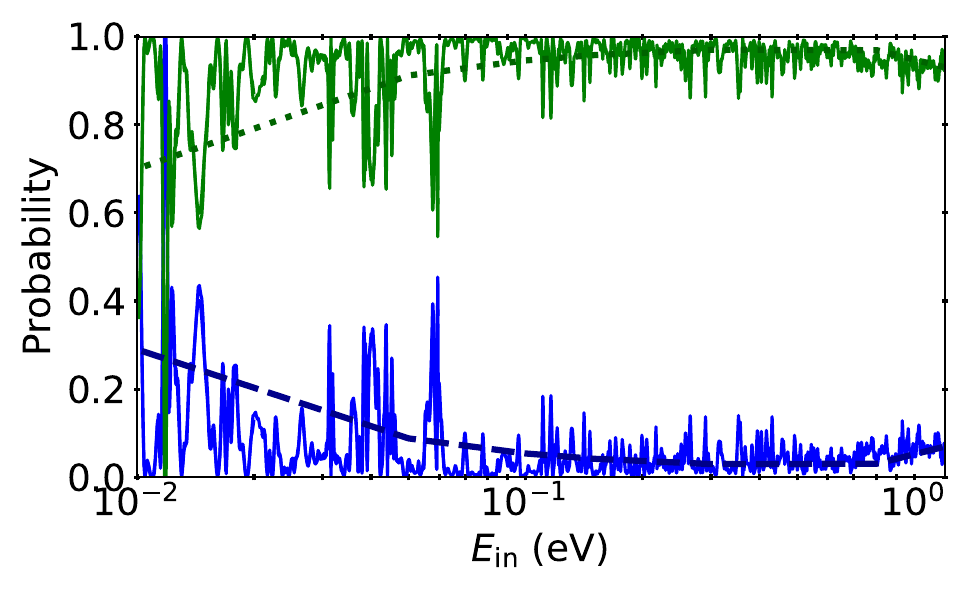}%
\caption{\label{fig:stickHvsener} Reflection (green) and absorption (blue) probabilities for H atoms scattering at W(110) surface with normal incidence as a function of incident energy $E_{\rm in}$. Dashed (solid) lines correspond to classical (quantum) results.}
\end{figure}

First, we investigate the absorption and reflection probabilities for hydrogen atoms scattering at the W(110) surface. We consider the range of incident energies $E_{\rm in}\in [0.01,1.2]$~eV that goes from the low energy regime, where quantum effects are expected, to the intermediate one that overlaps with the one {\color{teal}of interest for catalysis applications.}
We perform both classical and quantum dynamical calculations in order to characterize the regime in which quantum effects become important and identify a limit to such regime in terms of $E_{\rm in }$, hereinafter, the classical limit. 
Above the classical limit, classical dynamics can be safely used to describe the dynamics of the system. 
We restrict our study to normal incidence conditions. Furthermore, in the present work dissipation of energy is not considered and thus surface adsorption can only occur as a transient phenomena. 
In turn, this means that for converged calculations in terms of propagation time, the only two possible exit channels are reflection and bulk absorption. 
Figure~\ref{fig:stickHvsener} depicts the computed absorption ($P_{\rm abs}$) and reflection ($P_{\rm ref}$) probabilities as a function of energy. Classical (quantum) results are shown in dotted (solid) lines. 
Since $P_{\rm ref}= 1 - P_{\rm abs}$, only the absorption results are commented. 

The classical results reveal three regimes in terms of $E_{\rm in}$. 
Below $E_{\rm in} = 200$ meV, the absorption probability decreases as $E_{\rm in}$ increases. 
Between  $E_{\rm in }= 200$~meV and $1$~eV, it remains essentially constant, after which it increases for $E_{\rm in}>1$ eV. 
To understand the low $E_{\rm in}$ behavior, it is necessary to consider the absorption mechanism itself. 
For an incident atom to penetrate into the bulk region and contribute to the absorption probability, it must transfer part of its initial energy from the $Z$ degree of freedom, $E_{z}$, to the parallel components ($X$ and $Y$). This process is illustrated by the minimum energy path shown in Fig.~\ref{fig:potZcuts}(b).
This occurs naturally on a corrugated PES where trajectories are continuously deflected as they approach the surface, leading to a transient dynamical trapping.
From this transient state atoms may either be reflected back into the gas phase or penetrate into the bulk. 
In the former case,  atoms in the transient trapping state, need to transfer sufficient energy back into the $Z$ degree of freedom to escape from the surface.
For small values of $E_{\rm in}$, the transient dynamical adsorption state has a long lifetime, as the atom needs to recover almost all its kinetic energy in the $Z$ degree of freedom to escape, something that is less likely to occur at lower energies. Consequently, the lower the incident energy, the longer the atom remains trapped and the more time it has to explore the PES and eventually find a pathway into the bulk.
This leads to an increase in the absorption probability as $E_{\rm in}$ decreases.
By contrast, for $E_{\rm in}> 1$ eV,  $P_{\rm abs}$ increases with the incident energy, as atoms have enough energy to overcome the energy barriers that are present at the surface. 
In the intermediate regime the two effects compete: the lifetime of the dynamical trapped state decreases with $E_{\rm in}$ while the ability to overcome the PES barriers increases. This competition gives rise to the region of almost constant absorption probability in terms of $E_{\rm in}$ ($P_{\rm abs}\approx 5\%$).

Our results qualitatively match those of Ref.~\onlinecite{Yang2025} for large values of the incident energy. However, they disagree for lower values of $E_{\rm in}$. The discrepancy arises because the simulations in Ref.~\onlinecite{Yang2025} are stopped at $t=1$~ps, which is too short for $E_{\rm in}<1$~eV when energy dissipation is neglected, as in the present case.
It is also interesting to compare our classical dynamics results for the W(110) surface with those presented in Ref.~\onlinecite{Becerra2020} for the W(100). 
The W(100) surface is rougher compared to W(110), which is a close-packed surface. On the more open W(100) surface, absorption increases as the energy decreases below 200~meV. For instance, at $E_{\rm in} = 100$~meV, $P_{\rm abs}^{\rm W(100)} \approx 0.35$, which is $3.5$ times larger than the corresponding value for the W(110) surface.

Fig.~\ref{fig:stickHvsener} also shows quantum dynamical predictions, obtained with the MCTDH method, for the absorption probability as a function of $E_{\rm in }$. 
At odds with the classical curve that is a smooth function of $E_{\rm in}$, the quantum results exhibit oscillations arising from several mechanisms: interference,~\cite{Moix2009} diffraction,~\cite{Medina2008} and, specially, the emergence of resonances due to the presence of bound states at the surface~\cite{Medina2008,Bortolani1986,Sanz2007}. 
For incident energies exceeding 150~meV, quantum and classical results become closer, consistent with a transition to the classical regime.
For the largest energies studied here, however, small resonances persist, introducing minor discrepancies between the classical and quantum predictions for certain values of $E_{\rm in}$.
Eventually, for incident energies much larger than the characteristic energy scale of the surface bound states, the quantum curve would smooth to the classical one. 
Note, however, that the potential wells are approximately $3$~eV deep (see Fig.~\ref{fig:potZcuts}) and  reaching incident energies well above this value is computationally expensive because it requires the use of a denser primitive grid (and a large SPF basis) to represent high momentum states. Such calculations are beyond the scope of this work.
More interestingly, for $E_{\rm in}< 100$ meV the discrepancies between classical and quantum results are larger. 
In this region, two features allow to distinguish the classical and quantum predictions. 
First, there is an overall overestimation of $P_{\rm abs}$ by the classical approach. As commented before, in order to have a finite $P_{\rm abs}$, it is necessary  to transfer energy to the $X$ and $Y$ degrees of freedom when the atom is at the surface. However, the density of available states in the quantum case is discrete, meaning that less final states are available for this purpose compared to the classical picture.
This gives rise to a reduced quantum absorption probability, as a larger portion of the wave-packet is directly back-scattered. This point is further disused together with the diffraction patterns in section~\ref{sec:res-diff}. 
The second remarkable feature of the quantum absorption curve is the emergence of sharp resonances for $E_{\rm in}< 100$~meV that can enhance the absorption probability to values above 0.40 and even close to 1 as $E_{\rm in}$ decreases. Note that the incident energies scale is logarithmic, and thus, the low energy resonances that seem broader in the plot are indeed narrower than those at high values of $E_{\rm in}$.

Identifying precisely the origin of all the resonances that appear in the absorption curve is a nearly impossible task, as different mechanism overlap giving rise to its complex structure. 
However, it is possible to identify the dominant mechanisms responsible for the resonances in certain intervals of $E_{\rm in}$.
Resonances in elastic scattering emerge due to processes that transfer energy from the incident degree of freedom to the remaining ones. 
In the case of atomic scattering on a rigid surface, as considered here, this energy transfer, $\Delta E$, can only occur from the normal to the parallel degrees of freedom of the incident atom.~\cite{Bortolani1986}
For a given mechanism leading to a value of $\Delta E$, it must be accomplished $\Delta E = E_{\rm in}+ |E_b^{j}|$, with $E_b^{j}$ the energy of the $j$-th bound state at the surface. That means that the energy transfer between the perpendicular and parallel degrees of freedom has to coincide with the sum of the incident energy plus the binding energy of a given state $j$ at the surface. 
In our case, the potential for H on W(110) is deep, with a large density of bound states below $E=0$, hence allowing for a large amount of possible resonances. 
Consequently each of the mechanisms discussed below gives rise to signals in the interval of incident energies $(0,\Delta E)$ which correspond to coupling with bound states with binding energies $E_b\in(-\Delta E, 0)$. 
In the context of semi-classical theory, resonances emerge as a consequence of the existence of periodic allowed orbits.~\cite{Bonnet2022}

In this respect, SAR mechanisms have been described in the bibliography. In what respects to our system, there are two SAR mechanisms that can give rise to resonances at a rigid surface: the diffraction-mediated SAR (DMSAR)~\cite{Sanz2007,Bortolani1986,Medina2008,Sanz2007} and the focused SAR.~\cite{Sanz2007,Miret2001,Benedeck1983}
To disentangle these mechanisms, we follow previous works in assuming that the intermediate scattering state to which the incident wave-packet couples can be approximately described in separable form, that is, $\Psi(X,Y,Z)\simeq f(X)g(Y)h(Z)$.
Note that the actual wave function of the system requires of a large sum of product forms to be represented (see Subsec. \ref{sec:method-Q}). Here we use this picture only as an analysis tool to identify the origin of the resonance structures appearing in the absorption curve.

\begin{figure}
\includegraphics[width=1.0\linewidth]{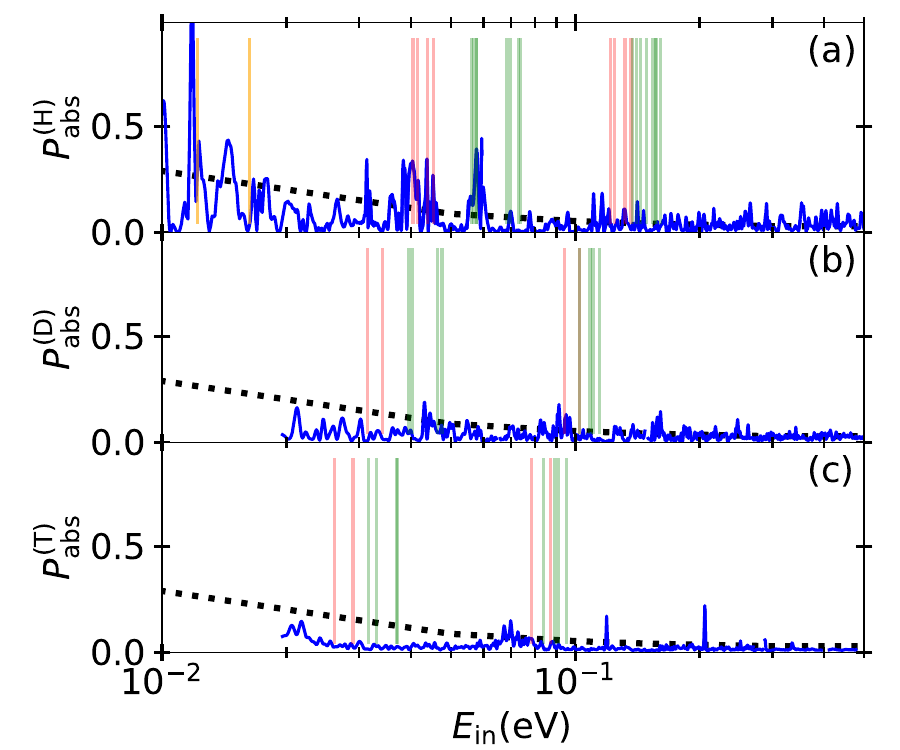} 
\caption{\label{fig:stickisotopes} 
Absorption probability as a function of the incident energy computed with MCTDH (solid blue line) and classical (dotted black line) methods for (a) hydrogen, (b) deuterium and (c) tritium isotopes. Vertical lines mark the energy transfer to ${\rm XY}$ degrees of freedom due to DMSAR (gold) and focused SAR (red). We further include processes combining DMSAR and focused SAR mechanisms (green).}
\end{figure}

In the DMSAR mechanism, the  energy transfer between the $Z$ and $\{X,Y\}$ degrees of freedom of the incident wave-packet occurs through a diffraction process. The first order diffraction processes for H impinging on W(110) corresponds to $\Delta E\approx$~12 (16)~meV for the $X$ ($Y$) directions.
Fig.~\ref{fig:stickisotopes}(a) depicts the absorption curve of hydrogen on W(110) for $E_{\rm in }$<~400~meV. The golden vertical lines mark the energy of the two lowest allowed diffraction processes. Based on the previous discussion of the classical results, these are likely responsible for the resonance structure that is observed for $E_{\rm in }\lesssim$~20~ meV.
While for the DMSAR mechanism a corrugated potential is the only ingredient needed, a more spectacular resonance effect can occur, called focused SAR or focused sticking,~\cite{Bortolani1986,Sanz2007}  if the potential supports also traversal bound states.  In this case,  it is possible to interchange energy between the perpendicular and transversal degrees of freedom of the incident atom at channels that are forbidden by diffraction, when bound states are present in the surface for the transversal degrees of freedom.
In Fig.~\ref{fig:stickisotopes}, panels (a)-(c), red vertical lines mark the energies corresponding to  bound states of H/D/T atoms at the hollow adsorption position of the W(110). For increasing energy, they correspond to the zero point energies  and excitations with one vibrational quantum in the $\{X,Y\}$ degrees of freedom (see appendix~\ref{app:vib}, for more details). Starting with H [see panel (a)], the energies of its vibrational spectra at W(110) surface help to explain the resonances appearing around values of $E_{\rm in}\sim 40$ meV and $\sim 125$~meV. Here we do not use the equal sign as the actual resonances arise from the coupling to the eigenstates of the full Hamiltonian. Finally, processes involving DMSAR in one of the degrees of freedom and focused SAR in the other one are also possible. In the figure, the energy of such processes is marked by green lines and help to explain the resonances that appear around $E_{\rm in }= 70$~meV (high intensity) and 140~meV (weak intensity). Note that only transversal bound states for the most stable position of H at the surface have  been considered. However, the PES supports more transversal bound states, e.g. localized at SB and LB high-symmetry positions, that add complexity to the resonance structure that is observed in that range of energies.

The above-mentioned discrepancies between the classical and quantum predictions arise due to the small mass of hydrogen. One may wonder, thus, whether they will persist if one considers the heavier isotopes of hydrogen.
Moreover, if the  resonance mechanisms described above are correct, they should also appear for D and T. Note that  the diffraction energies for D and T scale as $m_{\rm H}/m_i $, while the vibrational ones scale approximately with $\sqrt{m_{\rm H}/m_i}$ (see appendix~\ref{app:vib}), with $i= {\rm D, T}$. 
In Fig.~\ref{fig:stickisotopes} the absorption probability as a function of $E_{\rm in}$ for H, D and T is shown in panels (a), (b) and (c), respectively.
The first thing that one can observe is that, due to the larger density of states at the surface for D and T,  they approach faster to the classical limit in terms of $E_{\rm in}$. This is reflected in a smoother curve for the higher incident energies.
The region where the QD results are similar to the classical behavior is displaced to lower values of $E_{\rm in}$ as mass increases. For the case of H, differences in the absorption probability between the classical and quantum methods larger than 10\% appear for $E_{\rm in}<$~180~meV.  For deuterium such large differences appear only for $E_{\rm in}<$~100~meV. Finally, for the case of T, clear deviations between the classical and quantum predictions appear only for values of the incident energy lower than 70~meV.  Further, the expected diffraction resonances for D and T appear at very low energies, where computing the absorption curve is particularly challenging.  On the other hand, focused SAR processes as well as their combination with DMSAR seem to compose the regions where resonances are stronger for both D and T.

Despite  the above mentioned analysis, it is worth noting, however, that discrepancies at very low $E_{\rm in}$ persist, with the classical prediction overestimating the absorption probability. This confirms, as alluded before, that it is more difficult to transfer energy between perpendicular and parallel degrees of freedom in the quantum case compared to the classical one, which is intimately related to the reduced density of states at the surface in the quantized case. This is quite evident for T, whose absorption probability in the region $E_{\rm in}\in [35,70]$~meV, where no resonances are seen or expected, and the classical prediction still deviates from the quantum one. The difficulty of transferring energy between the parallel and perpendicular degrees of freedom gives place to an enhanced back-scattering probability (in our normal incidence conditions), as will be discussed in section~\ref{sec:res-diff}. Before concluding, it is worth emphasizing that the quantum effects observed in the absorption curve arise primarily from the quantization of surface-bound states in combination with diffraction processes.

\subsection{Dynamical isotopic effects}
\label{sec:res-dyniso}

\begin{figure}
\includegraphics[width=1.0\linewidth]{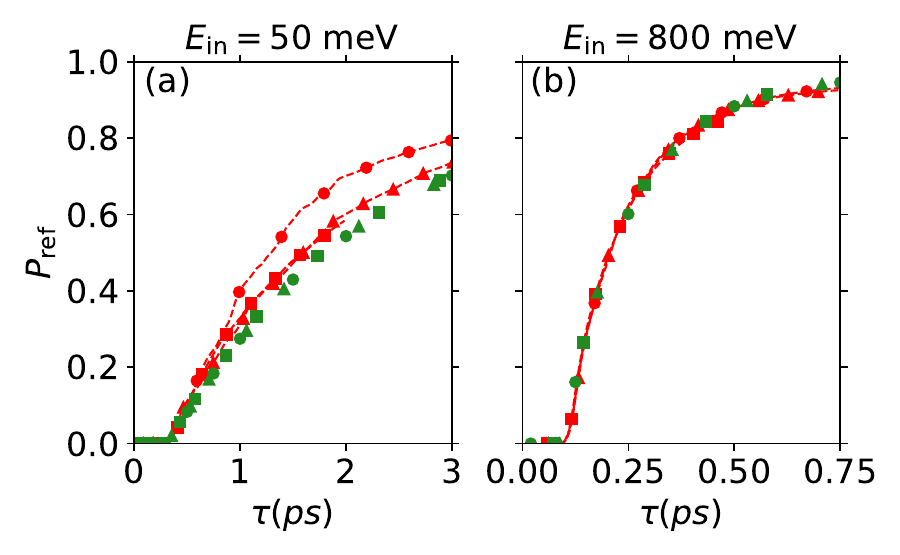} 
\caption{\label{fig:stickvstime} Time resolved cumulative reflection probability for hydrogen (circles), deuterium (triangles) and tritium (squares) computed with MCTDH (red) and classical (green) methods. The incident energy in (a) and (b) panels is 50 and 800 meV, respectively.
The time on the abscissa is scaled as $\tau = t\sqrt{m_{\rm H}/m_{\rm i}}$ with $i={\rm H, D, T}$. Lines are guides to the eye.}
\end{figure}

Further insight into isotopic effects is obtained by analyzing the time evolution of the scattering probability.
It is worth noting that, for an atom moving adiabatically in an external potential, as in our model, isotopic effects have a purely quantum origin.
In the classical picture, under the BOSS approximation, H, D, and T follow exactly the same trajectory in coordinate space if they start with the same initial conditions. The only difference is that heavier isotopes evolve more slowly in time.
The reason is that, when solving Newton's equation of motion for a particle moving in an external potential, the mass appears only as a scaling parameter of time. As a result, the dynamics of different isotopes become equivalent when expressed in terms of the scaled time $\tau=t\sqrt{m_{\rm H}/m_i}$, with $i={\rm H,D,T}$. In contrast, such a scaling is not possible in the quantum case, where the mass enters the Schr\"odinger equation through the kinetic-energy operator. Consequently, within a BOSS description such as the one employed here, isotopic effects arise exclusively from the quantum treatment. The classical absence of isotopic effects no longer holds in more realistic models that include surface motion or dissipative channels. In particular, coupling to phonons introduces mass-dependent energy transfer to the surface, while electronic friction may further differentiate the dynamics of H, D, and T. Figure~\ref{fig:stickvstime} confirms this scaling behavior. The classical reflection probability (green symbols), plotted as a function of the scaled time, collapses onto a single curve for H, D, and T at both low ($E_{\rm in}=50$~meV) and high ($E_{\rm in}=800$~meV) incident energies.

In contrast, the quantum description introduces isotope-dependent behavior through the mass dependence of the wave-packet propagation. Being a purely quantum effect, it is expected to be more important for low values of $E_{\rm in}$, consistent with the previous analysis of the absorption curves. In Fig.~\ref{fig:stickvstime}(a), the quantum evolution of $P_{\rm ref}$ in terms of the scaled time for $E_{\rm in}=50$~meV is shown. Results in this case are integrated over the energies of a wave-packet whose translational energy is $E_{\rm in}$. As it can be seen, isotopic dynamical effects appear for low values of the incident energy, which makes the dynamics of H different from that of D and T. These two latter isotopes exhibit behavior that is barely distinguishable between them and that it is close to the classical limit.
It is worth noticing that the isotopic effect on  $P_{\rm ref}(t\to\infty)$ (averaged over wave-packet energies) is small, even in the quantum case. 
Remarkably, for $E_{\rm in} = 800$~meV,  quantum signatures are suppressed and, thus, isotopic differences diminish, and the average evolution of the quantum system converges toward the classical behavior. This shows that our quantum dynamical calculations approach correctly to the classical limit, which also validates their quality.

\begin{figure*}
\includegraphics[width=1.0\linewidth]{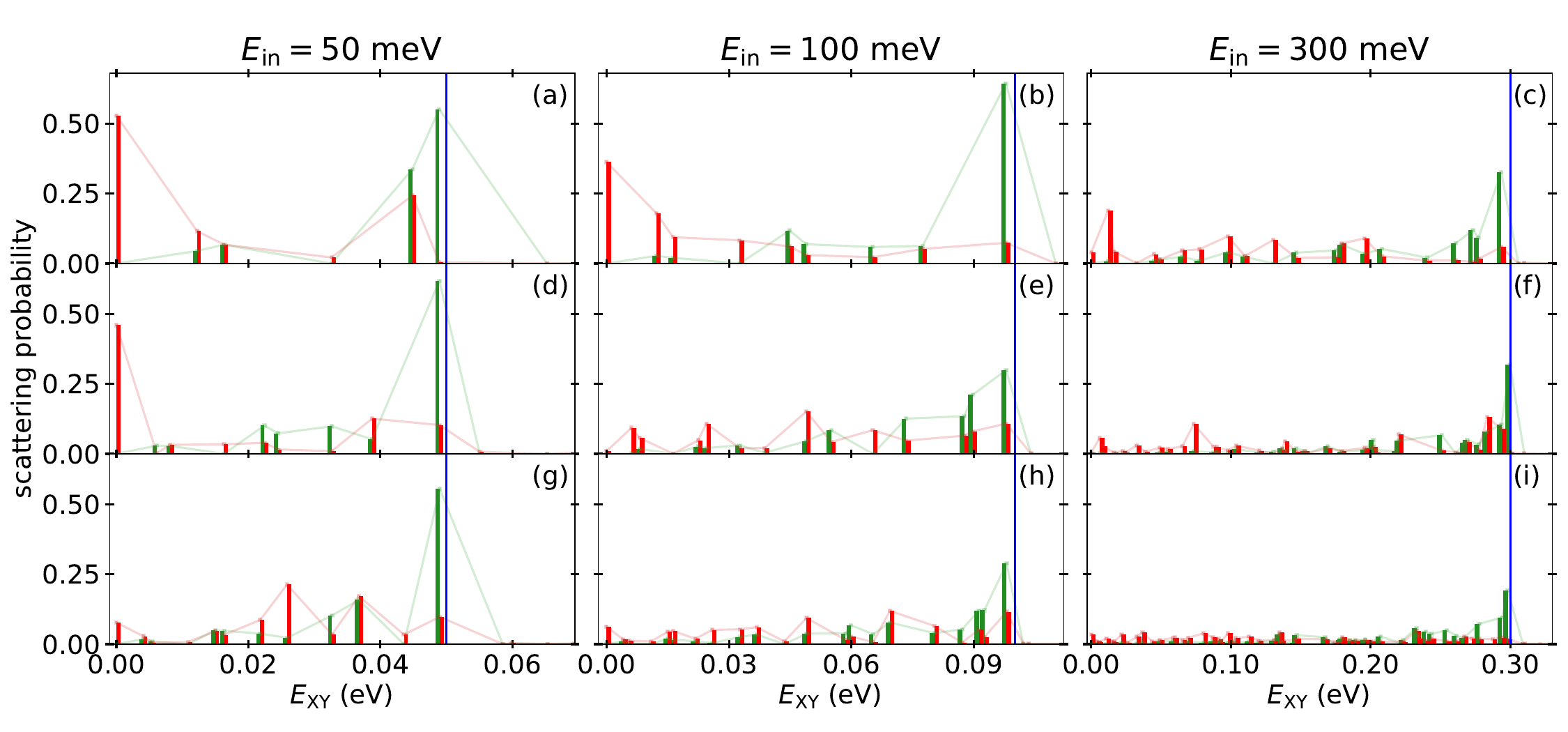} 
\caption{\label{fig:diffpatters} Energy resolved diffraction probability as a function of the energy transferred to the parallel (${\rm XY}$) degrees of freedom. Classical (quantum) results are shown as green (red) bars. Panels (a)-(c), (d)-(f), and (g)-(i) correspond to hydrogen, deuterium, and tritium isotopes, respectively. The blue vertical lines indicate the values of the incident energy labeled at the top of each column. Since diffraction channels are discrete, the lines serve only as visual guides.}
\end{figure*}

\subsection{Diffraction patterns}
\label{sec:res-diff}

Additional information about the origin of the quantum and isotopic effects that appear for low values of $E_{\rm in}$ is obtained from the analysis of the diffraction patterns in the reflection channel.
Classically, the diffraction patterns are evaluated by introducing Gaussian binning to the reflected states in the reciprocal space, with a Gaussian weight centered at the values of the quantum-allowed diffraction channels.
In this sense, they are semi-classical.~\cite{Rodriguez2019}
Although the semi-classical diffraction patterns already account for the quantization of the final states, the observed rainbow structures arise from the classical focusing of trajectories at specific exit angles induced by the corrugation of the PES. However, they do not account for the interference between different scattering pathways. The quantum diffraction patterns, which include these interference effects, are obtained by projecting the flux crossing the reflection CAP onto the corresponding diffraction states at a given energy.

Fig.~\ref{fig:diffpatters} shows the probabilities associated with the diffraction energy channels of energy $E_{\rm XY}$.
The semi-classical prediction for H, D, and T as a function of $E_{\rm XY}$ is reported for three different values of the incident energies: specifically, $E_{\rm in} = 50, 100, 300$~meV. Note that although no isotopic effects are present for the reflection probability in the classical picture, they do appear in the semi-classical diffraction patterns. The latter is evident when comparing the diffraction patterns for the three isotopes at the same $E_{\rm in}$.
This arises due to the quantization of the final diffraction states that depend themselves on the isotope mass.
Despite this consideration, the semi-classical diffraction patterns are qualitatively similar in the range of energies that we consider and for the three isotopes. In all cases, diffraction energy channels are populated up to the maximum allowed value of energy transfer to the parallel components, that is $E_{\rm XY} < E_{\rm in}$. Classical trajectories deflect continuously as they approach the corrugated potential induced by the W(110) surface; this gives rise to a negligible reflection probability for back-scattering events ($E_{\rm XY} = 0$), with the most probable exit channel being that with the maximum allowed value of $E_{\rm XY}$.

The comparison of the above mentioned semi-classical results with the quantum dynamical ones reveals the importance of quantum effects in this observable.
In the same figure we also show the quantum prediction for the diffraction patterns at energy $E_{\rm in}$, obtained from simulations with wave-packets centered at the three different values of $E_{\rm in}$.
For hydrogen scattering at W(110) with small $E_{\rm in}$, semiclassical and quantum diffraction patterns look quite different [see panel (a)]. 
Results show that the most probable quantum reflection channel is the back-scattering one. While classical trajectories can be continuously deflected, in the quantum case energy transfer to the in-plane degrees of freedom requires coupling to states with momentum parallel to the surface. Those states are discretized, at odds with what happens in the classical case, which makes the energy transfer less efficient. As the incident energy increases, the relative population of the high-energy diffraction channels with respect to the back-scattering channel increases. For $E_{\rm in} = 300$~meV, the back-scattering channel is no longer the most probable one and a quantum diffraction rainbow is observed.~\cite{Miret2012} Yet, the differences between the semi-classical and quantum predictions persist. 

With respect to isotopic effects, quantum diffraction channels are qualitatively different for the three isotopes. For deuterium, for the lowest energy ($E_{\rm in} = 50$~meV), back-scattering is still the most probable channel, but lower than for H, and the channel whose energy is closer to  $E_{\rm in}$ is enhanced with respect to H. In this sense D behaves  more classically than H. This approximation to the classical limit is clearer as the incident energy increases. Finally, in the case of tritium, the back-scattering channel is highly suppressed even for the lowest $E_{\rm in}$ considered. On a final note, it is worth remarking that although the semi-classical approximation does not match the prediction of the quantum rainbow diffraction pattern,~\cite{Miret2012} some features of the classical patterns remain in the quantum ones. Notably, in the classical picture, some diffraction channels have vanishing probability that arises from geometric focusing effects induced by the corrugation of the PES. This is the case of the channel at 34~meV in panel Fig.~\ref{fig:diffpatters}(a), whose vanishing value would be in principle from geometric origin and not due to destructive interference.

\section{\label{sec:conclusions} Summary}
 In this work, we have investigated the quantum dynamics of H, D, and T scattering from W(110) by performing time-dependent wave-packet calculations within the MCTDH framework. The quantum results have been systematically compared with classical trajectory calculations on the same potential energy surface. To isolate the role of nuclear quantum effects, both approaches were carried out within a frozen-surface Born--Oppenheimer static surface model, in which energy dissipation through phonons and electron-hole pair excitations is neglected. This simplified model provides a well-defined reference for identifying quantum nuclear effects before incorporating dissipative energy channels in more realistic descriptions.

For hydrogen, the quantum and classical predictions of the absorption probability as a function of the incident energy are in reasonably good agreement for $E_{\rm in} > 200$~meV. At lower incident energies, the quantum dynamical treatment predicts that the absorption curve is dominated by resonances. Consequently, the classical approach is unable to reproduce the quantum results.
We show that the origin of these resonances can qualitatively be explained in terms of selective adsorption resonances and diffraction-mediated selective adsorption resonances. This interpretation is further supported by comparing the absorption curves for the three hydrogen isotopes. Our results confirm that the classical limit shifts toward smaller values of $E_{\rm in}$ as the isotope mass increases. Furthermore, the positions at which resonant structures appear in the sticking curve satisfy the expected mass-dependent relationships of their underlying diffraction and vibrational origin. Despite the persistence of quantum effects in the absorption probability computed at long propagation times, we demonstrate that isotopic effects of genuine quantum mechanical origin can still be identified and that they introduce differences in the dynamics of the system.
Finally, analysis of the diffraction patterns of reflected trajectories give further insight into the origin of such quantum effects. 
The classical treatment underestimates the back-scattering probability, that is the dominant channel at low incident energies within the quantum description. This behavior is intimately related to the quantization of  states at the surface as well as interference effects. The importance of back scattering in the quantum case, explains the reduced adsorption probability predicted by the quantum method compared with the classical one.
At intermediate values of the incident energy ($E_{\rm in}=300$~meV), both quantum and semi-classical methods predict the appearance of a quantum diffraction rainbow pattern for all three isotopes; however, semi-classical results still depart from the quantum ones.

\begin{acknowledgments}
We acknowledge funding funding from ADAGIO (Advanced Manufacturing Research Fellowship Programme in the Basque – New Aquitaine Region) from the European Union’s Horizon 2020 research and innovation programme under the Marie Sklodowska  Curie cofund Grant Agreement No. 101034379. O.G. acknowledges financial support
by the Gobierno Vasco-UPV/EHU Project No. IT1569-22 and
by the Spanish MCIN/AEI/10.13039/501100011033 (Grant No. PID2022-140163NB-I00).
This work was conducted in the scope of the transborder joint Laboratory QuantumChemPhys: Theoretical Chemistry and Physics at the Quantum Scale (ANR-10-IDEX-03-02).
Computer time was provided by the P\^ole Modélisation HPC facility of the Institut des Sciences Moléculaires, UMR 5255, CNRS, Université de Bordeaux, cofunded by the Nouvelle Aquitaine region as well as the MCIA (Mésocentre de Calcul Intensif Aquitain).
\end{acknowledgments}

\section*{Data Availability Statement}

The data and plotting scripts to generate the results presented in this work are publicly available in the repository~\cite{Bombin2026zenodo} at https://doi.org/10.5281/zenodo.21134823

\appendix

\section{Vibrational spectra at the surface}
\label{app:vib}

\begin{figure}
    \centering
    \includegraphics[width=1.0\linewidth]{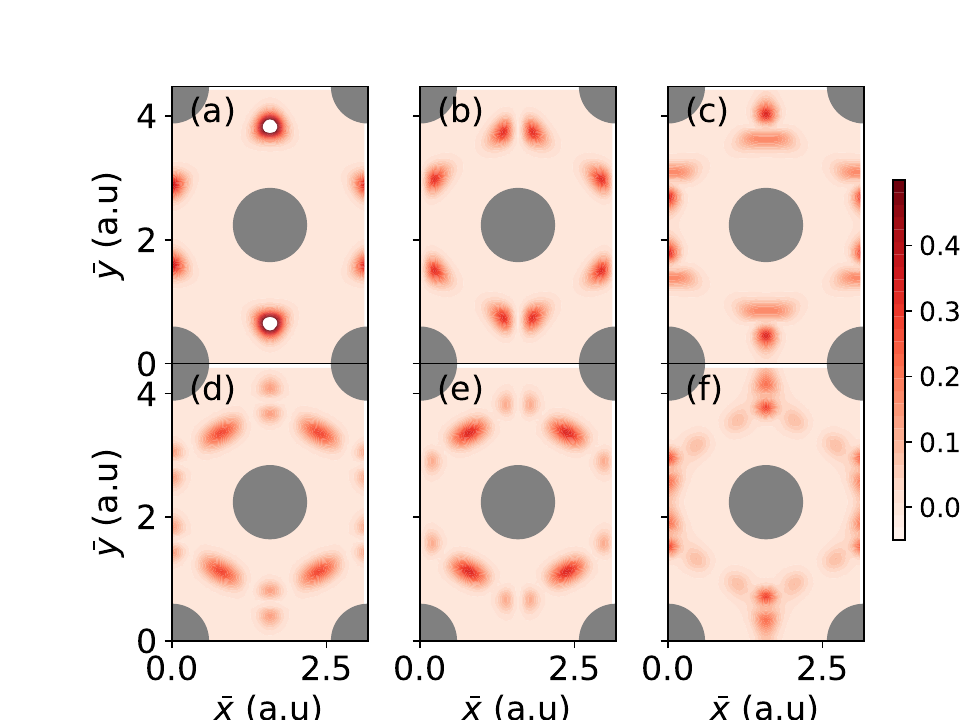}
    \caption{Cuts of the H nuclear probability density at ${\rm Z}=1.1$~{\AA} for the (a) ground state of H adsorbed at the W(110) surface and for (b)-(f) vibrational states 1, 3, 5, 6 and 7 of Table~\ref{table:boundstates}.}
    \label{fig:boundstates}
\end{figure}

We evaluate the bound states of H isotopes on the W(110) surfaces's primitive cell, making use of the improve relaxation algorithm. 
Table~\ref{table:boundstates} depicts the vibrational energies of H isotopes adsorbed a the W(110) surface. For each isotope, the energy reference is taken as the one of its ground state. Those ground states for H, D and T have zero point energies (ZPE) of  175~meV, 124~meV and 102~meV, respectively. 
Notice that these energies satisfy the expected relation for an harmonic potential, $E_{\rm ZPE}^i=E_{\rm ZPE}^{\rm H}\sqrt{m_{\rm H}/m_i}$ with $i = {\rm D, T}$. 
The corresponding ground state density for H, centered at Hollow position, is depicted in Fig.~\ref{fig:boundstates}(a). In the primitive cell, two hollow positions are present, and, thus the ground states are doubly degenerated.

The first excitations in the system correspond to vibrational excitations in $x$ and $y$ degrees of freedom. In the case of H, two quasi-degenerate excitations appear for each of the degrees of freedom, corresponding to symmetric, antisymetric wave functions. As the density in such quasi-degenerated states is barely distinguishable, in Fig~\ref{fig:boundstates} we only show the density corresponding to the symmetric cases [see panels (b) and (c)] for the H case. The small splitting between this states arises due to the delocalization of H at hollow position, that allows it to explore the anharmonic region of the well even for these low energy excitations.
With our numerical precision, we consider two states to be degenerated if their difference is less than 1~meV. With this criteria, for the cases of D and T, these first vibrational excitations are doubly degenerated. This is in agreement with the fact that D and T states are more localized and thus the spectra becomes more classical.
Notice that this first vibrational excitations for the three isotopes do not satisfy the relation $E_{\rm vib}^i=E_{\rm vib}^{\rm H}\sqrt{m_{\rm H}/m_i}$ with $i = {\rm D, T}$ that is expected for the states in an harmonic oscillator potential. This tells us that PES  at the W(110) surface becomes anharmonic for energies larger than 250~meV.

For higher excitation energies, the density of states increases rapidly and the SB and LB positions start to be populated as $E_{vib}>100$~meV. On the contrary, the repulsive atop positions remain unpopulated at these low energies. The density of some representative H states in this range of energy is illustrated in panels (d) - (f), Fig.~\ref{fig:boundstates}. As it can be seen, as energy grows, first bridge positions are populated, and subsequently a nodal structure appears. The latter signaling vibrational excitations at that bridge positions also.

\begin{table}[tb]
    \caption{Excited state number $n$, and its energy, $E_{\rm vib}$, for the first vibrational bound-states of H, D and T at the W(110) surface.  The numbers in parenthesis  are the corresponding degeneracies. }
    \label{table:boundstates}
    \begin{ruledtabular}
    \begin{tabular}{c ccc }
         state number & $E_{\rm vib}^{\rm H}$(meV) & $E_{\rm vib}^{\rm D}$(meV) & $E_{\rm vib}^{\rm T}$(meV) \\ \hline
            0  & 0 (2)   & 0 (2) & 0 (2)\\
            1  &  81 (1) & 63 (2) & 52 (2)  \\
            2  &  83 (1)& 69 (2) & 58 (2) \\
            3  &  88 (1)& 98 (2) & 82 (2)  \\
            4  &  91 (1)&99 (2) & 98 (2)  \\
            5  &  114 (1)&108 (1)    & 102(2) \\
            6  &  118 (1)&109 (1)   & 115 (2)\\
            7  &  134 (1)&114 (1)    & 116 (2) \\
            8  &  136 (1)&117 (1)     & 128 (1) \\
            9  &  147 (1)& 126 (1)   &  129 (1)\\
            10 &  149 (1)& 130 (1)    & 134 (1) 
    \end{tabular}
    \end{ruledtabular}
\end{table}

\section{Convergence of the vibrational states at the surface}
\label{app:convvib}
\begin{figure}[tb!]
\begin{center}
		\includegraphics[width=1.00\linewidth]{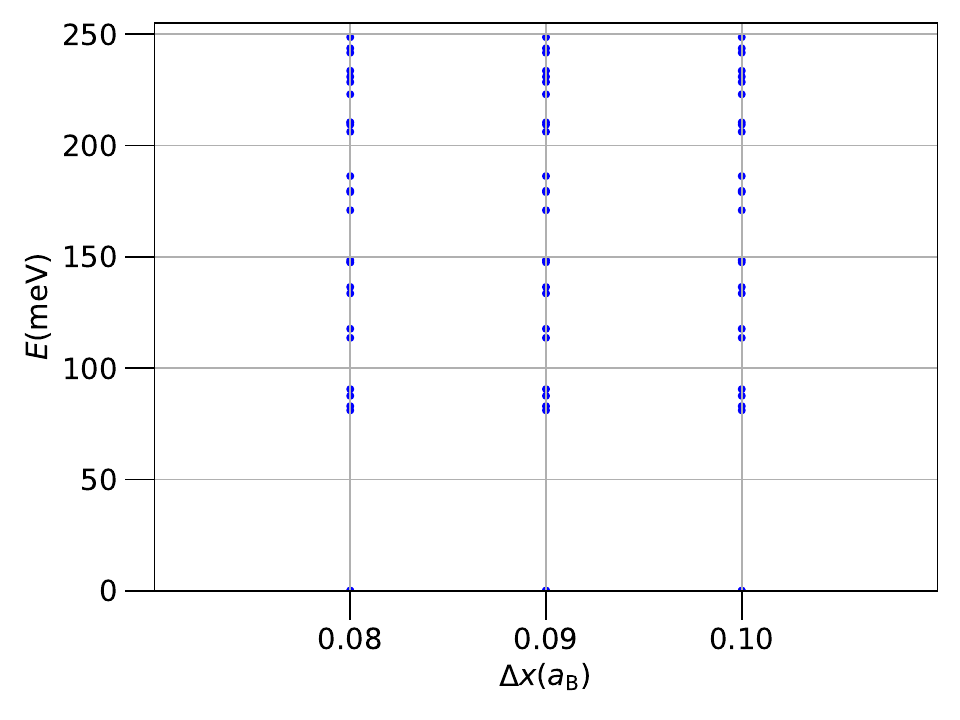}
	\caption{\label{fig:vibstatesconv}Vibrational states for H atom adsorbed at the W(110) surface computed for different grid spacing and converged in terms of SPFs. }
\end{center}
\end{figure}

A key point in the analysis of the resonances that appear in the absorption curve of Fig.~\ref{fig:stickisotopes} is the evaluation of the vibrational states of hydrogen isotopes at the tungsten surface. The computed vibrational spectra can be sensitive to the grid where the TDWP calculations are done and the representation of the potential on it. To check the convergence of the vibrational spectra with the $xy$-grid spacing we have considered spacings $\Delta x = \Delta y = 0.08, 0.09$ and $0.10a_B$. The results depicted in Fig.~\ref{fig:vibstatesconv} show that the vibrational spectra is already converged for $\Delta x = \Delta y = 0.10a_B$.

\color{black}

\bibliography{HonW110MCTDH}

@article{Auerbach2021,
author = {Auerbach, Daniel J. and Tully, John C. and Wodtke, Alec M.},
title = {{Chemical dynamics from the gas-phase to surfaces}},
journal = {Nat. Sci.},
volume = {1},
number = {1},
pages = {e10005},
doi = {https://doi.org/10.1002/ntls.10005},
url = {https://onlinelibrary.wiley.com/doi/abs/10.1002/ntls.10005},
year = {2021}
}

@article{Barrios2021,
author = {Martin Barrios, Raidel and Galparsoro, Oihana and Martínez Mesa, Aliezer and Uranga-Pi{\~n}a, Llinersy and Crespos, C{\'e}dric and Larregaray, Pascal},
title = {{Translational Inelasticity of Hydrogen Atoms Scattering off Hydrogen-Covered W(110) Surfaces}},
journal = {J. Phys. Chem. C},
volume = {125},
number = {25},
pages = {14075-14081},
year = {2021},
doi = {10.1021/acs.jpcc.1c03211},
URL = {https://doi.org/10.1021/acs.jpcc.1c03211}
}

@Article{Barrios2022,
author ="Martin-Barrios, Raidel and Hertl, Nils and Galparsoro, Oihana and Kandratsenka, Alexander and Wodtke, Alec M. and Larrégaray, Pascal",
title  ={{H atom scattering from W(110): A benchmark for molecular dynamics with electronic friction.}},
journal  ="Phys. Chem. Chem. Phys.",
year  ="2022",
volume  ="24",
issue  ="35",
pages  ="20813-20819",
publisher  ="The Royal Society of Chemistry",
doi  ="10.1039/D2CP01850K",
url  ="http://dx.doi.org/10.1039/D2CP01850K"
}

@article{Barrios2024,
author = {Martin Barrios, Raidel and Omar, Norhan and Galparsoro, Oihana and Crespos, Cedric and Larregaray, Pascal},
title = {{Specular Scattering of H Atoms off Pristine and H-Covered Tungsten Surfaces}},
journal = {J. Phys. Chem. C},
volume = {128},
number = {31},
pages = {13333-13338},
year = {2024},
doi = {10.1021/acs.jpcc.4c03231},
URL = {https://doi.org/10.1021/acs.jpcc.4c03231}
}

@article{Becerra2020,
title = {{Atomic scattering of H and N on W(100): Effect of lattice vibration and electronic excitations on the dynamics}},
journal = {Surf. Sci},
volume = {701},
pages = {121678},
year = {2020},
issn = {0039-6028},
doi = {https://doi.org/10.1016/j.susc.2020.121678},
url = {https://www.sciencedirect.com/science/article/pii/S0039602820303691},
author = {C. Ibarguen Becerra and C. Crespos and O. Galparsoro and P. Larregaray}
}

@article{Beck2000,
	title = {{The multiconfiguration time-dependent Hartree (MCTDH) method: a highly efficient algorithm for propagating wavepackets}},
	journal = {Phys. Rep},
	volume = {324},
	number = {1},
	pages = {1-105},
	year = {2000},
	issn = {0370-1573},
	doi = {https://doi.org/10.1016/S0370-1573(99)00047-2},
	url = {https://www.sciencedirect.com/science/article/pii/S0370157399000472},
	author = {M.H. Beck and A. Jäckle and G.A. Worth and H.-D. Meyer}
}

@article{Benedeck1983,
  title = {{Experimental evidence for kinematical focusing in the inelastic scattering of helium from the NaF(001) surface}},
  author = {Benedek, G. and Brusdeylins, G. and Toennies, J. P. and Doak, R. B.},
  journal = {Phys. Rev. B},
  volume = {27},
  issue = {4},
  pages = {2488--2493},
  numpages = {0},
  year = {1983},
  month = {Feb},
  publisher = {American Physical Society},
  doi = {10.1103/PhysRevB.27.2488},
  url = {https://link.aps.org/doi/10.1103/PhysRevB.27.2488}
}

@article{Bi2024,
    author = {Bi, Rui-Hao and Dou, Wenjie},
    title = {{Electronic friction near metal surface: Incorporating nuclear quantum effect with ring polymer molecular dynamics}},
    journal = {J. Chem. Phys.},
    volume = {160},
    number = {7},
    pages = {074110},
    year = {2024},
    month = {02},
    issn = {0021-9606},
    doi = {10.1063/5.0187646},
    url = {https://doi.org/10.1063/5.0187646}
}

@article{Bocan2020,
  title = {{Anomalous KCl(001) Surface Corrugation from Fast He Diffraction at Very Grazing Incidence}},
  author = {Bocan, G. A. and Breiss, H. and Szilasi, S. and Momeni, A. and Casagrande, E. M. Staicu and Gravielle, M. S. and S\'anchez, E. A. and Khemliche, H.},
  journal = {Phys. Rev. Lett.},
  volume = {125},
  issue = {9},
  pages = {096101},
  numpages = {5},
  year = {2020},
  month = {Aug},
  publisher = {American Physical Society},
  doi = {10.1103/PhysRevLett.125.096101},
  url = {https://link.aps.org/doi/10.1103/PhysRevLett.125.096101}
}

@article{Bonfanti2015,
    author = {Bonfanti, Matteo and Jackson, Bret and Hughes, Keith H. and Burghardt, Irene and Martinazzo, Rocco},
    title = {{Quantum dynamics of hydrogen atoms on graphene. I. System-bath modeling}},
    journal = {J. Chem. Phys.},
    volume = {143},
    number = {12},
    pages = {124703},
    year = {2015},
    month = {09},
    issn = {0021-9606},
    doi = {10.1063/1.4931116},
    url = {https://doi.org/10.1063/1.4931116}
}

@article{Bonfanti2018,
doi = {10.1088/1361-648X/aac89f},
url = {https://doi.org/10.1088/1361-648X/aac89f},
year = {2018},
month = {jun},
publisher = {IOP Publishing},
volume = {30},
number = {28},
pages = {283002},
author = {Bonfanti, Matteo and Achilli, Simona and Martinazzo, Rocco},
title = {{Sticking of atomic hydrogen on graphene}},
journal = {J. Phys: Cond. Matt.}
}

@article{Bonnet2022,
    author = {Bonnet, L. and Crespos, C. and Monnerville, M.},
    title = {Chemical reaction thresholds according to classical-limit quantum dynamics},
    journal = {J. Chem. Phys.},
    volume = {157},
    number = {9},
    pages = {094114},
    year = {2022},
    month = {09},
    issn = {0021-9606},
    doi = {10.1063/5.0101311},
    url = {https://doi.org/10.1063/5.0101311}
}

@Article{Bortolani1986,
author={Bortolani, V.
and Levi, A. C.},
title={Atom-surface scattering theory},
journal={Riv. Nuovo Cimento},
year={1986},
month={Nov},
day={01},
volume={9},
number={11},
pages={1-77},
issn={1826-9850},
doi={10.1007/BF02724478},
url={https://doi.org/10.1007/BF02724478}
}

@article{Bridge2024,
    author = {Bridge, Oliver and Lazzaroni, Paolo and Martinazzo, Rocco and Rossi, Mariana and Althorpe, Stuart C. and Litman, Yair},
    title = {Quantum rates in dissipative systems with spatially varying friction},
    journal = {J. Chem. Phys.},
    volume = {161},
    number = {2},
    pages = {024110},
    year = {2024},
    month = {07},
    issn = {0021-9606},
    doi = {10.1063/5.0216823},
    url = {https://doi.org/10.1063/5.0216823}
}

@article{Bucalossi2024,
doi = {10.1088/1741-4326/ad64e5},
url = {https://doi.org/10.1088/1741-4326/ad64e5},
year = {2024},
month = {sep},
publisher = {IOP Publishing},
volume = {64},
number = {11},
pages = {112022},
author = {Bucalossi, J. and \textit{et al.}},
title = {{WEST full tungsten operation with an ITER grade divertor}},
journal = {Ncl. Fusion}
}

@article{Bunermann2015_Science,
author = {Oliver Bünermann  and Hongyan Jiang  and Yvonne Dorenkamp  and Alexander Kandratsenka  and Svenja M. Janke  and Daniel J. Auerbach  and Alec M. Wodtke },
title = {Electron-hole pair excitation determines the mechanism of hydrogen atom adsorption},
journal = {Science},
volume = {350},
number = {6266},
pages = {1346-1349},
year = {2015},
doi = {10.1126/science.aad4972}
}

@article{Bunermann2021_JPCA,
	author       = {Oliver B{\"u}nermann and Alexander Kandratsenka and Alec M. Wodtke},
	title        = {Inelastic scattering of {H} atoms from surfaces},
	journal      = {J. Phys. Chem. A},
	year         = {2021},
	volume       = {125},
	number       = {15},
	pages        = {3059--3076},
	doi          = {10.1021/acs.jpca.1c00361},
}

@article{Busnengo2000,
    author = {Busnengo, H. F. and Salin, A. and Dong, W.},
    title = {{Representation of the 6D potential energy surface for a diatomic molecule near a solid surface}},
    journal = {J. Chem. Phys.},
    volume = {112},
    number = {17},
    pages = {7641-7651},
    year = {2000},
    month = {05},
    issn = {0021-9606},
    doi = {10.1063/1.481377},
    url = {https://doi.org/10.1063/1.481377}
}

@article{Busnengo2008,
author = {Busnengo, H. Fabio and Martínez, Alejandra E.},
title = {{H$_2$ Chemisorption on W(100) and W(110) Surfaces}},
journal = {J. Phys. Chem. C},
volume = {112},
number = {14},
pages = {5579-5588},
year = {2008},
doi = {10.1021/jp711053c},
URL = {https://doi.org/10.1021/jp711053c}
}

@techreport{Campbell2024_ITER_IRP,
	author       = {D.~J. Campbell and A. Loarte and D. Boilson and X. Bonnin and P. de Vries and L. Giancarli and Y. Gribov and M. Henderson and S.~H. Kim and Ph. Lamalle and M. Lehnen and T. Luce and I. Nunes and A.~R. Polevoi and S.~D. Pinches and R.~A. Pitts and R. Reichle and M. Schneider and J. Snipes and J. van der Laan and G. Vayakis},
	title        = {{ITER Research Plan within the Staged Approach (Level {III} -- Final Version)}},
	institution  = {ITER Organization},
	year         = {2024},
	type         = {ITER Technical Report (ITR-24-005)},
	url          = {https://www.iter.org/sites/default/files/media/2024-04/itr-24-005-ok1.pdf}
}

@article{Chen2025,
    author = {Chen, Jingqi and Lee, Joonho and Dou, Wenjie},
    title = {{How to correct Ehrenfest nonadiabatic dynamics in open quantum systems: Ehrenfest plus random force $(E+\sigma)$ dynamics}},
    journal = {J. Chem. Phys.},
    volume = {162},
    number = {4},
    pages = {044104},
    year = {2025},
    month = {01},
    issn = {0021-9606},
    doi = {10.1063/5.0245114},
    url = {https://doi.org/10.1063/5.0245114}
}

@Article{Diaz2010,
author ="Díaz, C. and Olsen, R. A. and Auerbach, D. J. and Kroes, G. J.",
title  ="{{Six-dimensional dynamics study of reactive and non reactive scattering of H$_2$ from Cu(111) using a chemically accurate potential energy surface}}",
journal  ="Phys. Chem. Chem. Phys.",
year  ="2010",
volume  ="12",
issue  ="24",
pages  ="6499-6519",
publisher  ="The Royal Society of Chemistry",
doi  ="10.1039/C001956A",
url  ="http://dx.doi.org/10.1039/C001956A"
}

@article{Diaz2005,
  title = {{Quantum and classical dynamics of H$_2$ scattering from Pd(111) at off-normal incidence}},
  author = {D\'{i}az, C. and Somers, M. F. and Kroes, G. J. and Busnengo, H. F. and Salin, A. and Mart\'{i}n, F.},
  journal = {Phys. Rev. B},
  volume = {72},
  issue = {3},
  pages = {035401},
  numpages = {10},
  year = {2005},
  month = {Jul},
  publisher = {American Physical Society},
  doi = {10.1103/PhysRevB.72.035401},
  url = {https://link.aps.org/doi/10.1103/PhysRevB.72.035401}
}

@article{Dorenkamp2018,
    author = {Dorenkamp, Yvonne and Jiang, Hongyan and Köckert, Hansjochen and Hertl, Nils and Kammler, Marvin and Janke, Svenja M. and Kandratsenka, Alexander and Wodtke, Alec M. and Bünermann, Oliver},
    title = {{Hydrogen collisions with transition metal surfaces: Universal electronically nonadiabatic adsorption}},
    journal = {J. Chem. Phys.},
    volume = {148},
    number = {3},
    pages = {034706},
    year = {2018},
    month = {01},
    issn = {0021-9606},
    doi = {10.1063/1.5008982},
    url = {https://doi.org/10.1063/1.5008982}
}

@article{Ferrin2012,
title = {{Hydrogen adsorption, absorption and diffusion on and in transition metal surfaces: A DFT study}},
journal = {Surf. Sci},
volume = {606},
number = {7},
pages = {679-689},
year = {2012},
issn = {0039-6028},
doi = {https://doi.org/10.1016/j.susc.2011.12.017},
url = {https://www.sciencedirect.com/science/article/pii/S0039602811004985},
author = {Peter Ferrin and Shampa Kandoi and Anand Udaykumar Nilekar and Manos Mavrikakis}
}

@article{Gravielle2014,
  title = {{Semiquantum approach for fast atom diffraction: Solving the rainbow divergence}},
  author = {Gravielle, M. S. and Miraglia, J. E.},
  journal = {Phys. Rev. A},
  volume = {90},
  issue = {5},
  pages = {052718},
  numpages = {6},
  year = {2014},
  month = {Nov},
  publisher = {American Physical Society},
  doi = {10.1103/PhysRevA.90.052718},
  url = {https://link.aps.org/doi/10.1103/PhysRevA.90.052718}
}

@article{Gallo2024,
title = {{Wall conditions in WEST during operations with a new ITER grade, actively cooled divertor}},
journal = {Nucl. Mat. Energy},
volume = {41},
pages = {101741},
year = {2024},
issn = {2352-1791},
doi = {https://doi.org/10.1016/j.nme.2024.101741},
url = {https://www.sciencedirect.com/science/article/pii/S2352179124001649},
author = {A. Gallo and \textit{et al.}}
}

@Article{Galparsoro2016,
author ="Galparsoro, Oihana and Pétuya, Rémi and Busnengo, Fabio and Juaristi, Joseba Iñaki and Crespos, Cédric and Alducin, Maite and Larregaray, Pascal",
title  ="Hydrogen abstraction from metal surfaces: when electron–hole pair excitations strongly affect hot-atom recombination",
journal  ="Phys. Chem. Chem. Phys.",
year  ="2016",
volume  ="18",
issue  ="46",
pages  ="31378-31383",
publisher  ="The Royal Society of Chemistry",
doi  ="10.1039/C6CP06222A",
url  ="http://dx.doi.org/10.1039/C6CP06222A"
}

@article{Galparsoro2018,
    author = {Galparsoro, Oihana and Busnengo, H. Fabio and Martinez, Alejandra E. and Juaristi, Joseba Iñaki and Alducin, Maite and Larregaray, Pascal},
    title = {{Energy dissipation to tungsten surfaces upon hot-atom and Eley–Rideal recombination of H$_2$}},
    journal  ="Phys. Chem. Chem. Phys.",
    volume = {20},
    number = {33},
    pages = {21334-21344},
    year = {2018},
    month = {09},
    issn = {1463-9076},
    doi = {10.1039/c8cp03690j},
    url = {https://doi.org/10.1039/c8cp03690j}
}

@article{Henkelman2000,
    author = {Henkelman, Graeme and Uberuaga, Blas P. and Jónsson, Hannes},
    title = {A climbing image nudged elastic band method for finding saddle points and minimum energy paths},
    journal = {J. Chem. Phys.},
    volume = {113},
    number = {22},
    pages = {9901-9904},
    year = {2000},
    month = {12},
    issn = {0021-9606},
    doi = {10.1063/1.1329672},
    url = {https://doi.org/10.1063/1.1329672}
}

@article{Hertl2021,
author = {Hertl, Nils and Martin-Barrios, Raidel and Galparsoro, Oihana and Larr{\'e}garay, Pascal and Auerbach, Daniel J. and Schwarzer, Dirk and Wodtke, Alec M. and Kandratsenka, Alexander},
title = {{Random Force in Molecular Dynamics with Electronic Friction}},
journal = {J. Phys. Chem. C},
volume = {125},
number = {26},
pages = {14468-14473},
year = {2021},
doi = {10.1021/acs.jpcc.1c03436},
URL = {https://doi.org/10.1021/acs.jpcc.1c03436}
}

@Article{Hertl2022,
author ="Hertl, Nils and Kandratsenka, Alexander and Wodtke, Alec M.",
title  ="{Effective medium theory for bcc metals: electronically non-adiabatic H atom scattering in full dimensions}",
journal  ="Phys. Chem. Chem. Phys.",
year  ="2022",
volume  ="24",
issue  ="15",
pages  ="8738-8748",
publisher  ="The Royal Society of Chemistry",
doi  ="10.1039/D2CP00087C",
url  ="http://dx.doi.org/10.1039/D2CP00087C"
}

@article{Hodille2025,
title = {{Modelling fuel retention in the W divertor during the D/H/D changeover experiment in WEST}},
journal = {Nucl. Mat. Energy},
volume = {45},
pages = {101999},
year = {2025},
issn = {2352-1791},
doi = {https://doi.org/10.1016/j.nme.2025.101999},
url = {https://www.sciencedirect.com/science/article/pii/S2352179125001413},
author = {Hodille, E. A. and \textit{et al.}}
}

@article{Jackson1995,
    author = {Jackson, Bret and Persson, Mats},
    title = {{Effects of isotopic substitution on Eley–Rideal reactions and adsorbate‐mediated trapping}},
    journal = {J. Chem. Phys.},
    volume = {103},
    number = {14},
    pages = {6257-6269},
    year = {1995},
    month = {10},
    issn = {0021-9606},
    doi = {10.1063/1.470404},
    url = {https://doi.org/10.1063/1.470404}
}

@article{Jackle1996,
    author = {Jäckle, A. and Meyer, H.‐D.},
    title = {Product representation of potential energy surfaces},
    journal = {J. Chem. Phys.},
    volume = {104},
    number = {20},
    pages = {7974-7984},
    year = {1996},
    month = {05}
}

@article{Jiang2021,
author = {Jiang, Hongyan and Tao, Xuecheng and Kammler, Marvin and Ding, Feizhi and Wodtke, Alec M. and Kandratsenka, Alexander and Miller, Thomas F. III and B{\"u}nermann, Oliver},
title = {{Small Nuclear Quantum Effects in Scattering of H and D from Graphene}},
journal = {J. Phys. Chem. Lett..},
volume = {12},
number = {7},
pages = {1991-1996},
year = {2021},
doi = {10.1021/acs.jpclett.0c02933},
URL = {https://doi.org/10.1021/acs.jpclett.0c02933}
}

@article{Jin2024,
  title = {Analysis of lattice locations of deuterium in tungsten and its application for predicting deuterium trapping conditions},
  author = {Jin, Xin and Djurabekova, Flyura and Hodille, Etienne A. and Markelj, Sabina and Nordlund, Kai},
  journal = {Phys. Rev. Mater.},
  volume = {8},
  issue = {4},
  pages = {043604},
  numpages = {18},
  year = {2024},
  month = {Apr},
  publisher = {American Physical Society},
  doi = {10.1103/PhysRevMaterials.8.043604},
  url = {https://link.aps.org/doi/10.1103/PhysRevMaterials.8.043604}
}

@article{Kroes1999,
title = {{Six-dimensional quantum dynamics of dissociative chemisorption of H$_2$ on metal surfaces}},
journal = {Prog. Surf. Sci},
volume = {60},
number = {1},
pages = {1-85},
year = {1999},
issn = {0079-6816},
doi = {https://doi.org/10.1016/S0079-6816(99)00006-4},
url = {https://www.sciencedirect.com/science/article/pii/S0079681699000064},
author = {Geert-Jan Kroes}
}

@article{Lepetit2011,
    author = {Lepetit, Bruno and Lemoine, Didier and Medina, Zuleika and Jackson, Bret},
    title = {{Sticking and desorption of hydrogen on graphite: A comparative study of different models}},
    journal = {J. Chem. Phys.},
    volume = {134},
    number = {11},
    pages = {114705},
    year = {2011},
    month = {03},
    issn = {0021-9606},
    doi = {10.1063/1.3565446},
    url = {https://doi.org/10.1063/1.3565446}
}

@article{Maggi2024,
doi = {10.1088/1741-4326/ad3e16},
url = {https://doi.org/10.1088/1741-4326/ad3e16},
year = {2024},
month = {aug},
publisher = {IOP Publishing},
volume = {64},
number = {11},
pages = {112012},
author = {Maggi, {C.F} and \textit{et al.}},
title = {{Overview of T and D–T results in {JET} with {ITER}-like wall}},
journal = {Nucl.Fusion}
}

@article{Mandal2022,
    author = {Mandal, Souvik and Gatti, Fabien and Bindech, Oussama and Marquardt, Roberto and Tremblay, Jean Christophe},
title = {{Stochastic multi-configuration time-dependent Hartree for dissipative quantum dynamics with strong intramolecular coupling}},
journal = {J. Chem. Phys.},
volume = {157},
number = {14},
pages = {144105},
year = {2022},
month = {10},
issn = {0021-9606},
doi = {10.1063/5.0105308}
}

@article{Medina2008,
    author = {Medina, Zuleika and Jackson, Bret},
    title = {Quantum studies of light particle trapping, sticking, and desorption on metal and graphite surfaces},
    journal = {J. Chem. Phys.},
    volume = {128},
    number = {11},
    pages = {114704},
    year = {2008},
    month = {03},
    issn = {0021-9606},
    doi = {10.1063/1.2890043},
    url = {https://doi.org/10.1063/1.2890043}
}

@article{Meijer2001,
author = {Meijer, Anthony J. H. M. and Farebrother, Adam J. and Clary, David C. and Fisher, Andrew J.},
title = {{Time-Dependent Quantum Mechanical Calculations on the Formation of Molecular Hydrogen on a Graphite Surface via an Eley-Rideal Mechanism}},
journal = {J. Phys. Chem. A},
volume = {105},
number = {11},
pages = {2173-2182},
year = {2001},
doi = {10.1021/jp003839+},
URL = {https://doi.org/10.1021/jp003839+}
}

@article{mey06:179,
title = {{Calculation and selective population of vibrational levels with the Multiconfiguration Time-Dependent Hartree (MCTDH) algorithm}},
journal = {Chem. Phys.},
volume = {329},
number = {1},
pages = {179-192},
year = {2006},
issn = {0301-0104},
doi = {https://doi.org/10.1016/j.chemphys.2006.06.002},
url = {https://www.sciencedirect.com/science/article/pii/S0301010406003119},
author = {Hans-Dieter Meyer and Frédéric Le Quéré and Céline Léonard and Fabien Gatti}
}

@article{Miret2001,
  title = {Focused sticking of light mass particles in physisorption},
  author = {Miret-Art\'es, S. and Manson, J. R.},
  journal = {Phys. Rev. B},
  volume = {63},
  issue = {12},
  pages = {121404},
  numpages = {4},
  year = {2001},
  month = {Mar},
  publisher = {American Physical Society},
  doi = {10.1103/PhysRevB.63.121404},
  url = {https://link.aps.org/doi/10.1103/PhysRevB.63.121404}
}

@article{Miret2012,
title = {Classical theory of atom–surface scattering: The rainbow effect},
journal = {Surf. Sci. Reports},
volume = {67},
number = {7},
pages = {161-200},
year = {2012},
issn = {0167-5729},
doi = {https://doi.org/10.1016/j.surfrep.2012.03.001},
url = {https://www.sciencedirect.com/science/article/pii/S0167572912000179},
author = {Salvador Miret-Artés and Eli Pollak}
}

@article{Moix2009,
  title = {{Semiclassical initial-value-representation study of helium scattering from Cu(110)}},
  author = {Moix, Jeremy M. and Pollak, Eli},
  journal = {Phys. Rev. A},
  volume = {79},
  issue = {6},
  pages = {062507},
  numpages = {8},
  year = {2009},
  month = {Jun},
  publisher = {American Physical Society},
  doi = {10.1103/PhysRevA.79.062507},
  url = {https://link.aps.org/doi/10.1103/PhysRevA.79.062507}
}

@article{Muzas2024,
  title = {{Semiquantum versus quantum methods for grazing-incidence fast-atom diffraction: Influence of the wave-packet size}},
  author = {Muzas, A. S. and Frisco, L. and Bocan, G. A. and D\'iaz, C. and Gravielle, M. S.},
  journal = {Phys. Rev. A},
  volume = {109},
  issue = {4},
  pages = {042823},
  numpages = {12},
  year = {2024},
  month = {Apr},
  publisher = {American Physical Society},
  doi = {10.1103/PhysRevA.109.042823},
  url = {https://link.aps.org/doi/10.1103/PhysRevA.109.042823}
}

@article{Olsen2002,
    author = {Olsen, R. A. and Busnengo, H. F. and Salin, A. and Somers, M. F. and Kroes, G. J. and Baerends, E. J.},
    title = {{Constructing accurate potential energy surfaces for a diatomic molecule interacting with a solid surface: H$_2$+Pt(111) and H$_2$+Cu(100)}},
    journal = {J. Chem. Phys.},
    volume = {116},
    number = {9},
    pages = {3841-3855},
    year = {2002},
    month = {03},
    issn = {0021-9606},
    doi = {10.1063/1.1446852},
    url = {https://doi.org/10.1063/1.1446852}
}

@article{Omar2025,
author = {Omar, Norhan and Galparsoro, Oihana and Truflandier, Lionel and Crespos, Cedric and Larregaray, Pascal},
title = {{How Surface Reconstruction Affects Hydrogen Dissociation on the N-Precovered W(110): A Theoretical Perspective}},
journal = {J. Phys. Chem. C},
volume = {129},
number = {21},
pages = {9706-9716},
year = {2025},
doi = {10.1021/acs.jpcc.5c00732},
URL = {https://doi.org/10.1021/acs.jpcc.5c00732}
}

@article{Petuya2014,
author = {P{\'e}tuya, R. and Crespos, C. and Quintas-Sanchez, E. and Larr{\'e}garay, P.},
title = {{Comparative Theoretical Study of H$_2$ Eley–Rideal Recombination Dynamics on W(100) and W(110)}},
journal = {J. Phys. Chem. C},
volume = {118},
number = {22},
pages = {11704-11710},
year = {2014},
doi = {10.1021/jp501679n},
URL = {https://doi.org/10.1021/jp501679n}
}

@article{Persson1995,
    author = {Persson, Mats and Jackson, Bret},
    title = {{Flat surface study of the Eley–Rideal dynamics of recombinative desorption of hydrogen on a metal surface}},
    journal = {J. Chem. Phys.},
    volume = {102},
    number = {2},
    pages = {1078-1093},
    year = {1995},
    month = {01},
    issn = {0021-9606},
    doi = {10.1063/1.469456},
    url = {https://doi.org/10.1063/1.469456}
}

@article{Pijper2000,
    author = {Pijper, E. and Kroes, G. J. and Olsen, R. A. and Baerends, E. J.},
    title = {{The effect of corrugation on the quantum dynamics of dissociative and diffractive scattering of H$_2$ from Pt(111)}},
    journal = {J. Chem. Phys.},
    volume = {113},
    number = {18},
    pages = {8300-8312},
    year = {2000},
    month = {11},
    issn = {0021-9606},
    doi = {10.1063/1.1314377},
    url = {https://doi.org/10.1063/1.1314377}
}

@article{Pijper2001,
title = {{Six-dimensional quantum dynamics of scattering of (v=0, j=0) H$_2$ from Pt(111): comparison to experiment and to classical dynamics results}},
journal = {Chem. Phys. Lett.},
volume = {347},
number = {4},
pages = {277-284},
year = {2001},
issn = {0009-2614},
doi = {https://doi.org/10.1016/S0009-2614(01)01074-0},
url = {https://www.sciencedirect.com/science/article/pii/S0009261401010740},
author = {E Pijper and M.F Somers and G.J Kroes and R.A Olsen and E.J Baerends and H.F Busnengo and A Salin and D Lemoine}
}

@article{Preston2025,
author = {Preston, Riley J. and Ke, Yaling and Rudge, Samuel L. and Hertl, Nils and Borrelli, Raffaele and Maurer, Reinhard J. and Thoss, Michael},
title = {{Nonadiabatic Quantum Dynamics of Molecules Scattering from Metal Surfaces}},
journal = {J. Chem. Theory Comput.},
volume = {21},
number = {3},
pages = {1054-1063},
year = {2025},
doi = {10.1021/acs.jctc.4c01586},
URL = {https://doi.org/10.1021/acs.jctc.4c01586}
}

@article{Rodriguez2019,
author = {Rodríguez-Fernández, A. and Bonnet, L. and Crespos, C. and Larr{\'e}garay, P. and Díez Mui{\~n}o, R.},
title = {{When Classical Trajectories Get to Quantum Accuracy: The Scattering of H$_2$ on Pd(111)}},
journal = {J. Phys. Chem. Lett.},
volume = {10},
number = {24},
pages = {7629-7635},
year = {2019},
doi = {10.1021/acs.jpclett.9b02742},
URL = {https://doi.org/10.1021/acs.jpclett.9b02742}
}

@article{Rodriguez2021,
    author = {Rodríguez-Fernández, Alberto and Bonnet, Laurent and Larrégaray, Pascal and Muiño, Ricardo Díez},
    title = {Ab initio molecular dynamics of hydrogen on tungsten surfaces},
    journal = {Phys. Chem. Chem. Phys.},
    volume = {23},
    number = {13},
    pages = {7919-7925},
    year = {2021},
    month = {04},
    issn = {1463-9076},
    doi = {10.1039/d0cp05423b},
    url = {https://doi.org/10.1039/d0cp05423b}
}

@article{Rodriguez2023,
author = {Rodríguez-Fernández, A. and Bonnet, L. and Larr{\'e}garay, P. and Díez Mui{\~n}o, R.},
title = {{How Adsorbed Oxygen Atoms Inhibit Hydrogen Dissociation on Tungsten Surfaces}},
journal = {J. Phys. Chem. Lett.},
volume = {14},
number = {5},
pages = {1246-1252},
year = {2023},
doi = {10.1021/acs.jpclett.2c03684},
URL = {https://doi.org/10.1021/acs.jpclett.2c03684}
}

@article{Sanz2007,
title = {{Selective adsorption resonances: Quantum and stochastic approaches}},
journal = {Phys. Rep.},
volume = {451},
number = {2},
pages = {37-154},
year = {2007},
issn = {0370-1573},
doi = {https://doi.org/10.1016/j.physrep.2007.08.001},
url = {https://www.sciencedirect.com/science/article/pii/S0370157307003250},
author = {A.S. Sanz and S. Miret-Artés},
}

@article{Sha2002,
    author = {Sha, Xianwei and Jackson, Bret and Lemoine, Didier},
    title = {{Quantum studies of Eley–Rideal reactions between H atoms on a graphite surface}},
    journal = {J. Chem. Phys.},
    volume = {116},
    number = {16},
    pages = {7158-7169},
    year = {2002},
    month = {04},
    issn = {0021-9606},
    doi = {10.1063/1.1463399},
    url = {https://doi.org/10.1063/1.1463399}
}

@article{Shi2025,
author = {Shi, Lei and Schr{\"o}der, Markus and Meyer, Hans-Dieter and Peláez, Daniel and Wodtke, Alec M. and Golibrzuch, Kai and Sch{\"o}nemann, Anna-Maria and Kandratsenka, Alexander and Gatti, Fabien},
title = {{Full Quantum Dynamics Study for H Atom Scattering from Graphene}},
journal = {J. Phys. Chem. A},
volume = {129},
number = {7},
pages = {1896-1907},
year = {2025},
doi = {10.1021/acs.jpca.4c06712},
URL = {https://doi.org/10.1021/acs.jpca.4c06712}
}

@article{Shi2023,
    author = {Shi, Lei and Schröder, Markus and Meyer, Hans-Dieter and Peláez, Daniel and Wodtke, Alec M. and Golibrzuch, Kai and Schönemann, Anna-Maria and Kandratsenka, Alexander and Gatti, Fabien},
    title = {{Quantum and classical molecular dynamics for H atom scattering from graphene}},
    journal = {J. Chem. Phys.},
    volume = {159},
    number = {19},
    pages = {194102},
    year = {2023},
    month = {11},
    issn = {0021-9606},
    doi = {10.1063/5.0176655},
    url = {https://doi.org/10.1063/5.0176655}
}

@article{Stark2023,
author = {Stark, Wojciech G. and Westermayr, Julia and Douglas-Gallardo, Oscar A. and Gardner, James and Habershon, Scott and Maurer, Reinhard J.},
title = {{Machine Learning Interatomic Potentials for Reactive Hydrogen Dynamics at Metal Surfaces Based on Iterative Refinement of Reaction Probabilities}},
journal = {J. Phys. Chem. C},
volume = {127},
number = {50},
pages = {24168-24182},
year = {2023},
doi = {10.1021/acs.jpcc.3c06648},
URL = {https://doi.org/10.1021/acs.jpcc.3c06648}
}

@article{Song2022,
author = {Song, Qingfei and Zhang, Xingyu and Gatti, Fabien and Miao, Zekai and Zhang, Qiuyu and Meng, Qingyong},
title = {{Multilayer Multiconfiguration Time-Dependent Hartree Study on the Mode-/Bond-Specific Quantum Dynamics of Water Dissociation on Cu(111)}},
journal = {J. Phys. Chem. A},
volume = {126},
number = {36},
pages = {6047-6058},
year = {2022},
doi = {10.1021/acs.jpca.2c03092},
URL = {https://doi.org/10.1021/acs.jpca.2c03092}
}

@article{Viaud2024,
author = {Viaud, L. T. and Ibarguen Becerra, C. and Crespos, C. and Bonnet, L. and Larregaray, P.},
title = {{Improved Theoretical Description of the H$_2$ Chemisorption Dynamics on the W(100) Surface}},
journal = {J. Phys. Chem. C},
volume = {128},
number = {41},
pages = {17410-17417},
year = {2024},
doi = {10.1021/acs.jpcc.4c04679},
URL = {https://doi.org/10.1021/acs.jpcc.4c04679}
}

@article{Viaud2025,
    author = {Viaud, L. T. and Somers, M. and Crespos, C. and Bonnet, L. and Larregaray, P.},
    title = {{Classical dynamics in a quantum spirit: Refining semi-classical corrections for the scattering of H$_2$ on W(100)}},
    journal = {J. Chem. Phys.},
    volume = {163},
    number = {5},
    pages = {054114},
    year = {2025},
    month = {08},
    issn = {0021-9606},
    doi = {10.1063/5.0272407},
    url = {https://doi.org/10.1063/5.0272407}
}

@Article{Xiong2024,
	author ="Xiong, Longlong and Zhang, Liang and Zhao, Bin and Jiang, Bin",
	title  ="{{Six-dimensional quantum dynamics of an Eley–Rideal reaction between gaseous and adsorbed hydrogen atoms on Cu(111)}}",
	journal  ="Faraday Discuss.",
	year  ="2024",
	volume  ="251",
	issue  ="0",
	pages  ="457-470",
	publisher  ="The Royal Society of Chemistry",
	doi  ="10.1039/D3FD00163F",
	url  ="http://dx.doi.org/10.1039/D3FD00163F"
}

@article{Xiong2024JPCC,
author = {Xiong, Longlong and Jiang, Bin},
title = {{Comparison of Six-Dimensional Quantum and Quasi-Classical Dynamics of the Eley–Rideal Reaction of H(D) Atoms with D(H)-Covered Cu(111)}},
journal = {J. Phys. Chem. C},
volume = {128},
number = {22},
pages = {9003-9010},
year = {2024},
doi = {10.1021/acs.jpcc.4c01571},
URL = {https://doi.org/10.1021/acs.jpcc.4c01571}
}

@article{Yang2025,
title = {{Sticking, reflection, and abstraction behavior of hydrogen irradiated on (110) tungsten surfaces at 0.1-100 eV by molecular dynamics simulations using a machine learning potential}},
journal = {Acta Mater.},
volume = {297},
pages = {121306},
year = {2025},
issn = {1359-6454},
doi = {https://doi.org/10.1016/j.actamat.2025.121306},
url = {https://www.sciencedirect.com/science/article/pii/S1359645425005920},
author = {Sojeong Yang and Seungyun Kim and Takuji Oda}
}

@article{Zhou2021,
author = {Zhou, Xueyao and Zhang, Yaolong and Yin, Rongrong and Hu, Ce and Jiang, Bin},
title = {{Neural Network Representations for Studying Gas-Surface Reaction Dynamics: Beyond the Born-Oppenheimer Static Surface Approximation†}},
journal = {Chin. J. Chem..},
volume = {39},
number = {10},
pages = {2917-2930},
doi = {https://doi.org/10.1002/cjoc.202100303},
url = {https://onlinelibrary.wiley.com/doi/abs/10.1002/cjoc.202100303},
year = {2021}
}

@article{Bombin2026zenodo,
  author  = {Ra{\'u}l Bomb{\'\i}n and Oihana Galparsoro and Daniel Pel{\'a}ez and Jean-Christophe Tremblay and C{\'e}dric Crespos and Pascal Larregaray},
  title   = {{Data and plotting scripts for: Quantum isotope effects in the scattering of hydrogen isotopes from W(110)}},
  journal = {Zenodo},
  year    = {2026},
  volume  = {version: 1.0},
  doi     = {https://doi.org/10.5281/zenodo.21134823},
  url     = {https://doi.org/10.5281/zenodo.21134823}
}

\end{document}